\documentclass[preprint2]{emulateapj}
\shorttitle{Sandage-96 After the Explosion of SN~2004dj}
\shortauthors{Vink\'o et al.}

\begin{document}
\title{The Young, Massive, Star Cluster Sandage-96 After the
Explosion of SN~2004\lowercase{dj} in NGC~2403}
\author{J.~Vink\'o\altaffilmark{1,2,3}, K.~S\'arneczky\altaffilmark{1},
Z.~Balog\altaffilmark{4,1}, S.~Immler\altaffilmark{5},
B.~E.~K.~Sugerman\altaffilmark{6}, P.~J.~Brown\altaffilmark{7},
K.~Misselt\altaffilmark{4}, Gy.~M.~Szab\'o\altaffilmark{8},
Sz.~Csizmadia\altaffilmark{9}, M.~Kun\altaffilmark{10}, 
P.~Klagyivik\altaffilmark{11},
R.~J.~Foley\altaffilmark{12,13,14},
A.~V.~Filippenko\altaffilmark{12}, B.~Cs\'ak\altaffilmark{1}, and
L.~L.~Kiss\altaffilmark{15}}

\altaffiltext{1}{Department of Optics \& Quantum Electronics,
University of Szeged, D\'om t\'er 9, Szeged, H-6720 Hungary.}
\altaffiltext{2}{email vinko@physx.u-szeged.hu .}
\altaffiltext{3}{Department of Astronomy, University of Texas, 
Austin, TX 78712.}
\altaffiltext{4}{Steward Observatory, University of Arizona, Tucson,
AZ 85721.}
\altaffiltext{5}{Astrophysics Science Division, X-Ray Astrophysical
Laboratory Code 662, NASA Goddard Space Flight Center, Greenbelt, MD
20771.}
\altaffiltext{6}{Department of Physics and Astronomy, Goucher College,
Baltimore, MD 21204.}
\altaffiltext{7}{Department of Astronomy and Astrophysics, Penn State
University, 525 Davey Laboratory, University Park, PA 16802.}
\altaffiltext{8}{Department of Experimental Physics, University of
Szeged, Hungary.}
\altaffiltext{9}{Institut of Planetary Research, German Aerospace Center
D-12489 Berlin, Rutherfordstrasse 2, Germany.}
\altaffiltext{10}{Konkoly Observatory of Hungarian Academy of Sciences,
Budapest, Hungary.}
\altaffiltext{11}{Department of Astronomy, E\"otv\"os Lor\'and
University, Budapest, Hungary.}
\altaffiltext{12}{Department of Astronomy, University of California,
Berkeley, CA 94720-3411.}
\altaffiltext{13}{Center for Astrophysics, 60 Garden St., Cambridge,
MA  02138.}
\altaffiltext{14}{Clay Fellow.}
\altaffiltext{15}{Institute of Astronomy, School of Physics,
University of Sydney, Sydney, NSW 2006, Australia.}

\begin{abstract}
The bright Type II-plateau supernova (SN) 2004dj occurred within the
young, massive stellar cluster Sandage-96 in a spiral arm of NGC~2403.
New multi-wavelength observations obtained with several ground-based
and space-based telescopes are combined to study the radiation from
Sandage-96 after SN~2004dj faded away.  Sandage-96 started to dominate
the flux in the optical bands starting September 2006 ($\sim$800~d
after explosion). The optical fluxes are equal to the pre-explosion
ones within the observational uncertainties. An optical Keck spectrum
obtained $\sim$900~d after explosion shows the dominant blue continuum
from the cluster stars shortward of 6000~\AA\ as well as strong SN
nebular emission lines redward.  The integrated spectral energy
distribution (SED) of the cluster has been extended into the
ultraviolet region by archival {\it XMM-Newton} and new {\it Swift}
observations, and compared with theoretical models. The outer parts of
the cluster have been resolved by the {\it Hubble Space Telescope},
allowing the construction of a color-magnitude diagram. The fitting of
the cluster SED with theoretical isochrones results in cluster ages
distributed between 10 and 40 Myr, depending on the assumed
metallicity and the theoretical model family.  The isochrone fitting
of the color-magnitude diagrams indicates that the resolved part of
the cluster consists of stars having a bimodal age distribution: a
younger population at $\sim$10--16 Myr, and an older one at
$\sim$32--100 Myr.  The older population has an age distribution
similar to that of the other nearby field stars.  This may be
explained with the hypothesis that the outskirts of Sandage-96 are
contaminated by stars captured from the field during cluster
formation.  The young age of Sandage-96 and the comparison of its pre-
and post-explosion SEDs suggest $12 \lesssim M_{\rm prog} \lesssim
20$~M$_\odot$ as the most probable mass range for the progenitor of
SN~2004dj.  This is consistent with, but perhaps slightly higher than,
most of the other Type II-plateau SN progenitor masses determined so
far.
\end{abstract}


\section{Introduction}

The theory of stellar evolution predicts that massive stars ($M
\gtrsim 8~M_\odot$) end their lives as core-collapse supernovae
(CC~SNe, e.g., \citealt{woo02}). In particular, after the
main-sequence phase the most massive stars undergo heavy mass loss,
become stripped stellar cores and explode as Type Ib/c supernovae
(SNe~Ib/c; see \citealt{fil97} for a discussion of SN spectral
classification).  Stars close to the lower mass limit of CC are
thought to produce Type II-plateau SNe (SNe~II-P). Recent observations
to detect the progenitors of CC~SNe support this scenario.  Currently
there are 10 SNe II (1987A, 1993J, 1999ev, 2003gd, 2004A, 2004et,
2005cs, 2006ov, 2008bk; the last seven are Type II-P) whose
progenitors have been directly identified in pre-explosion images (see
\citealt{hen06, li07, mat08, leonard08, smartt08}, and references
therein), and the mass estimates are $M \lesssim 15$--20 ~M$_\odot$
for all of them. Moreover, upper mass limits were derived for a number
of other SNe~II from nondetections of their progenitors \citep{vand03,
ms05,leonard08}, and the highest upper limit was found to be $M \approx
20$~M$_\odot$. These observations have led to the conclusion that
SNe~II-P likely originate from ``low-mass'' massive progenitors with
$M \lesssim 20$~M$_\odot$ \citep{li06, li07, smartt08}, and the fate
of stars with $M \gtrsim 20$~M$_\odot$ may be a SN~Ib/c explosion.

On the other hand, the progenitors of SNe~Ib/c (even the brightest and
closest ones) have escaped direct detection so far \citep{cr07a}.  The
most promising candidate is SN~2007gr, which occurred in a compact,
massive stellar cluster in NGC~1058 that has been detected with the
{\it Hubble Space Telescope (HST)} prior to explosion \citep{cr07b}.

SN~2004dj, the closest ($\sim 3.5$ Mpc, \citealt{v06}) and one of the
brightest SNe since SN~1987A, was a Type~II-P event. It occurred
within a young, massive cluster, Sandage-96 (S96) in NGC~2403.
SN~2004dj has been extensively studied through multi-wavelength
observations; see \citet{v06} (hereafter Paper~I) for references. In
particular, several attempts were made to infer the mass of the
progenitor by comparing the pre-explosion magnitudes and colors of S96
with theoretical spectral energy distributions (SEDs) to determine the
age, hence the turnoff mass, of the cluster. These resulted in a range
of possible progenitor masses from $M \approx 12$~M$_\odot$ to $M
\gtrsim 20$~M$_\odot$ depending on the assumed metallicity and/or
reddening \citep{maiz04, wang05, v06}. However, all of these studies
suffered from the age-reddening and age-metallicity degeneracy
\citep{rebuz86}, because the available pre-explosion observations
covered only the optical and near-infrared (NIR) bands.

The main aim of this paper is to derive further constraints on the
progenitor mass of SN~2004dj from post-explosion observations of S96,
made after its reappearance from the dimming light of the SN. The
cluster has been successfully redetected both by our ground-based and
space-based observations, showing no significant change in its optical
light level with respect to the pre-explosion level.  We have extended
the wavelength coverage of the observed SED into the ultraviolet (UV)
with new {\it Swift}/UVOT observations. Moreover, the cluster has been
partially resolved by our new $HST$/ACS observations, which provides
an additional opportunity to infer age constraints on its stellar
population.

The paper is organized as follows. In \S 2 we present our new
multi-wavelength observations made from the ground and space.  The SED
of the cluster and its color-magnitude diagram are constructed and
fitted with theoretical model predictions in \S 3. We discuss the
results in \S 4 and present our conclusions in \S 5.

\begin{figure*}
\begin{center}
\epsscale{1.0}
\plottwo{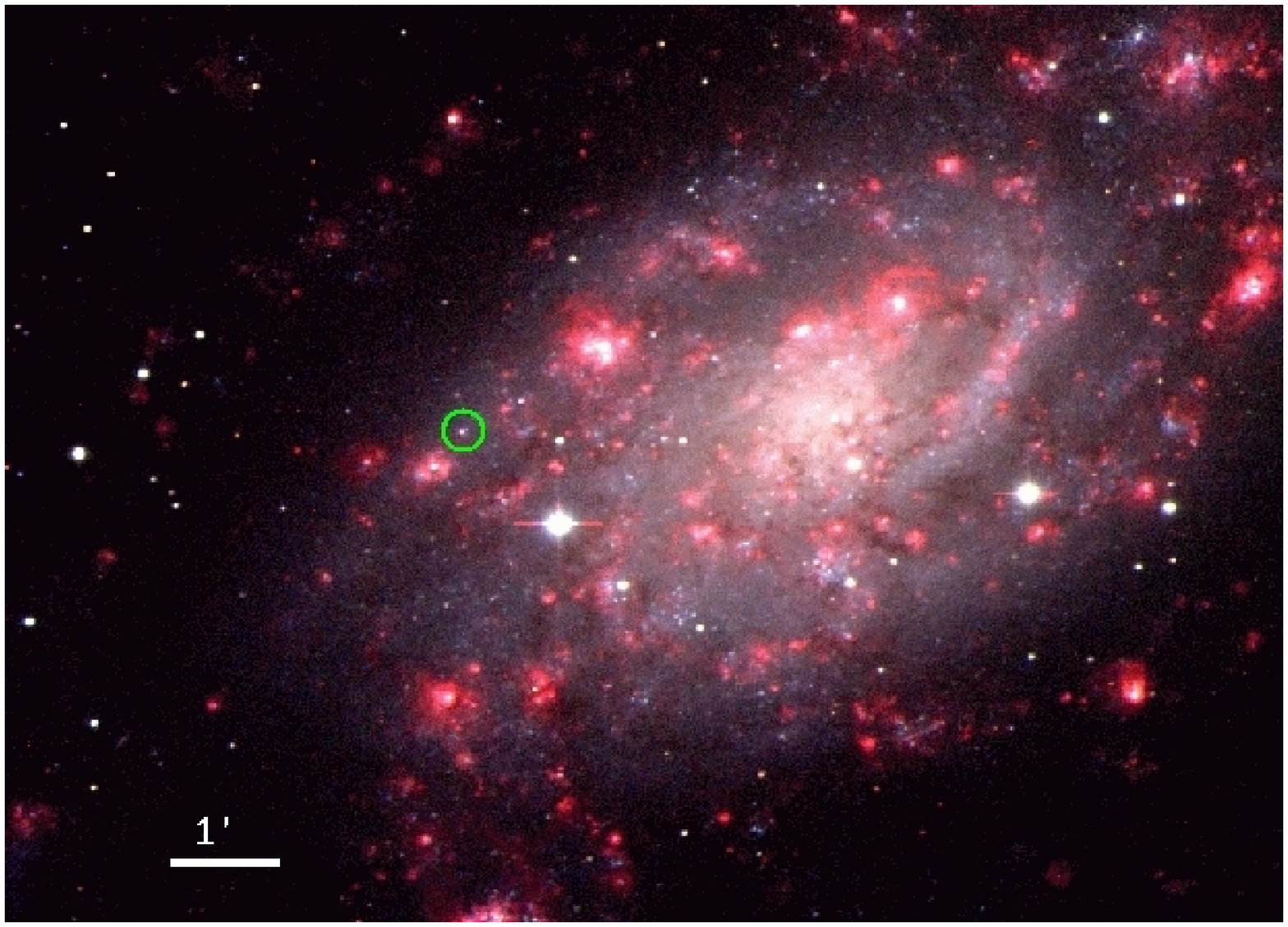}{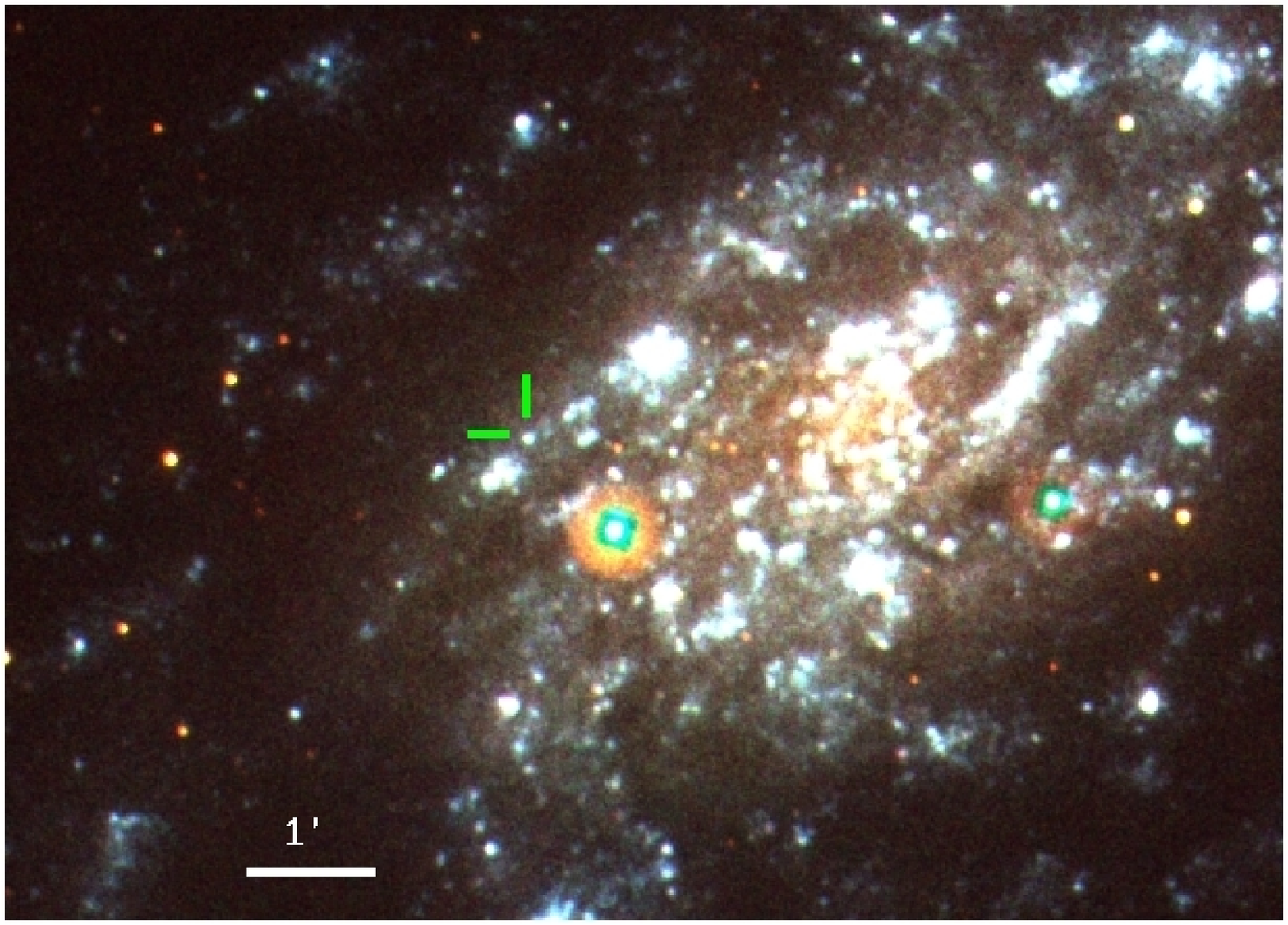}
\caption{Left panel: color-combined $B$, $V$, and H$\alpha$ frames of
NGC~2403 obtained with the 2.3~m Bok telescope at Steward Observatory
on 2007 Jan. 28 (UT dates are used throughout this paper).  Right
panel: {\it Swift}/UVOT image of NGC 2403 obtained on 2007 Dec. 2
($u$, uvw1, and uvw2 filters were selected as red, green, and blue
channels).  The field of view of both frames is $10.0' \times 7.2'$;
north is up and east is to the left.  The position of S96/SN~2004dj is
marked. }
\label{fig-bok}
\end{center}
\end{figure*}

\section{Observations}

\subsection{Optical Data}

\subsubsection{Ground-Based Photometry}

The light variation of SN~2004dj in the nebular phase was followed
from Konkoly Observatory (see Paper~I for a description of the
telescopes and detectors). In addition, Johnson $B$, Johnson $V$, and
narrow-band H$\alpha$ images were taken with the 90Prime camera on the
2.3~m Bok telescope at the Steward Observatory, Arizona
(Fig.~\ref{fig-bok}, left panel).

The magnitudes of SN~2004dj were calculated via aperture photometry
based on the same sequence of local standard stars as in Paper~I.  The
photometric data obtained after 2005 May are summarized in
Table~\ref{tbl-phot}.
 
\begin{figure*}
\epsscale{1.0}
\plottwo{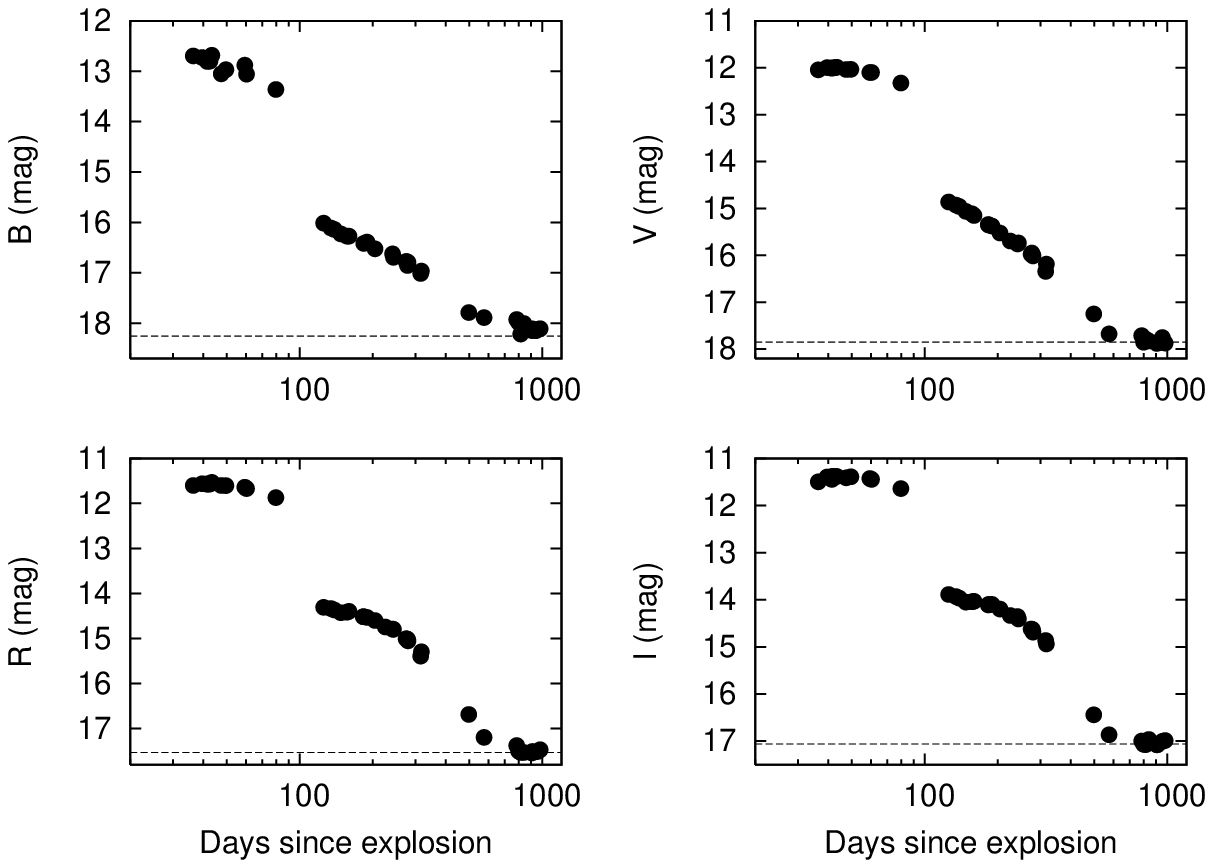}{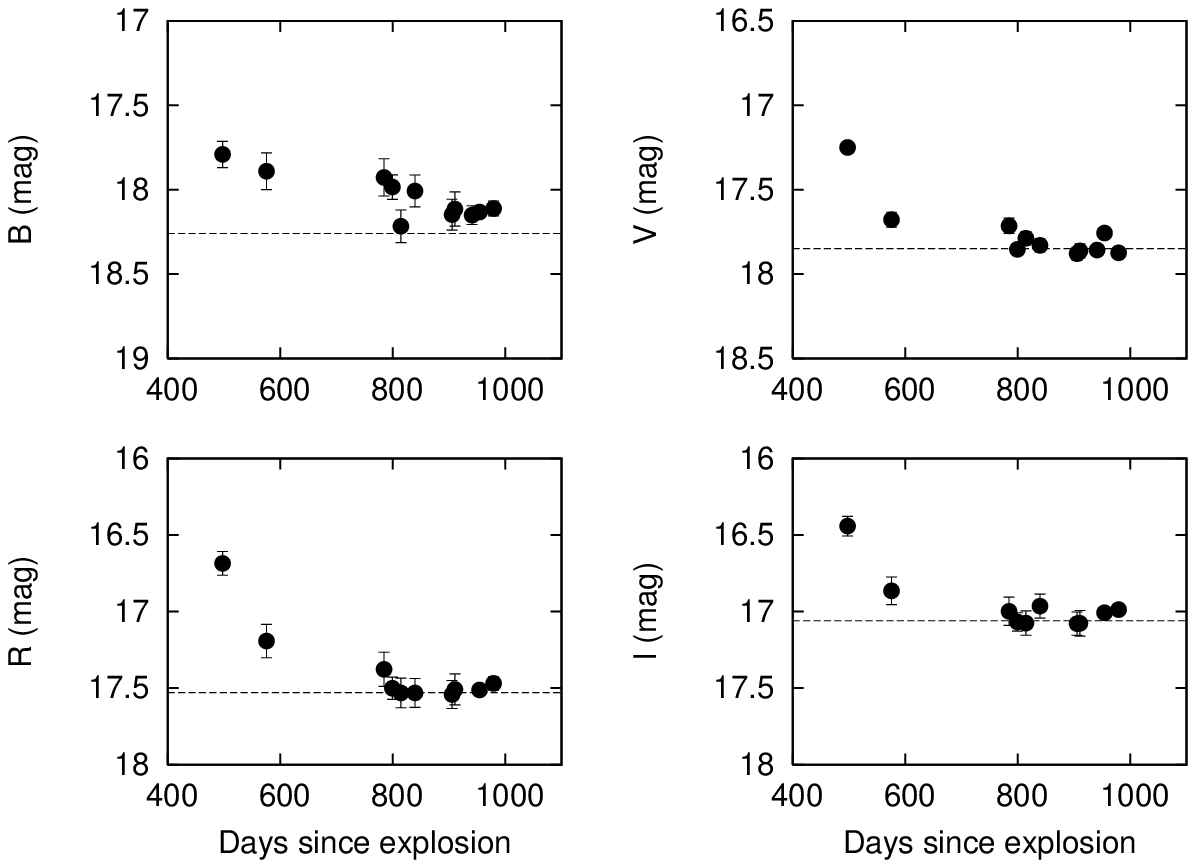}
\caption{$BVRI$ light curves of SN 2004dj from ground-based photometry. The 
horizontal lines mark the pre-explosion magnitudes of S96. In the left panel 
the scaling on the abscissa is logarithmic. The two right-hand panels show 
the same data as the left-hand ones, but focus on the region around 800~d.}
\label{fig-latelc}
\end{figure*}

The light curves are plotted in Figure~\ref{fig-latelc}.  Following
the usual decline in the nebular phase (starting $\sim 100$~d after
explosion), the light curves approached a constant level around day
800; see the two right-hand panels of Figure~\ref{fig-latelc}.

As expected, the flattening of the light curves is caused by the
increasing contribution of the radiation from S96, emerging from the
fading light of SN~2004dj.  In Figure~\ref{fig-latelc} the dotted
horizontal lines mark the pre-explosion magnitudes of S96 (see
Paper~I).

\begin{table*}
\begin{center}
\caption{Late-Time $BVRI$ Photometry of SN~2004dj. }
\label{tbl-phot}
\begin{tabular}{lccccccc}
\tableline\tableline
UT Date & JD $-$  & $t - t_{\rm expl}$ & $B$ & $V$ & $R$ & $I$ & Instrument \\
 & 2,450,000 & (d) & (mag) & (mag) & (mag) & (mag) & \\
\tableline
2005-11-09 & 3684.6 & 500 & 17.79 (0.08) & 17.25 (0.03) & 16.69 (0.08) & 16.44 (0.06) & Konkoly 0.6 m Schmidt \\ 
2006-01-27 & 3762.5 & 577 & 17.89 (0.11) & 17.68 (0.05) & 17.19 (0.11) & 16.86 (0.09) & Konkoly 0.6 m Schmidt \\ 
2006-08-23 & 3971.6 & 787 & 17.93 (0.11) & 17.71 (0.05) & 17.38 (0.11) & 17.00 (0.09) & Konkoly 0.6 m Schmidt \\ 
2006-09-07 & 3986.3 & 801 & 17.98 (0.07) & 17.85 (0.03) & 17.50 (0.07) & 17.07 (0.06) & Konkoly 1.0 m RCC \\ 
2006-09-22 & 4001.6 & 817 & 18.22 (0.10) & 17.79 (0.04) & 17.53 (0.10) & 17.08 (0.08) & Konkoly 0.6 m Schmidt \\ 
2006-10-17 & 4026.6 & 842 & 18.01 (0.09) & 17.83 (0.04) & 17.53 (0.09) & 16.96 (0.08) & Konkoly 0.6 m Schmidt \\ 
2006-12-22 & 4092.5 & 907 & 18.15 (0.09) & 17.88 (0.04) & 17.54 (0.09) & 17.08 (0.08) & Konkoly 0.6 m Schmidt \\ 
2006-12-27 & 4097.6 & 913 & 18.11 (0.10) & 17.86 (0.04) & 17.51 (0.10) & 17.08 (0.08) & Konkoly 0.6 m Schmidt \\ 
2007-01-28 & 4128.0 & 943 & 18.15 (0.06) & 17.86 (0.02)  &   --               &  --                 & Steward 2.3 m Bok \\
2007-02-09 & 4141.4 & 956 & 18.13 (0.08) & 17.76 (0.04) & 17.51 (0.06) & 17.01 (0.08) & Konkoly 1.0 m RCC \\ 
2007-03-06 & 4166.3 & 981 & 18.11 (0.08) & 17.87 (0.03) & 17.47 (0.06) & 16.99 (0.08) & Konkoly 1.0 m RCC \\ 
\tableline
\end{tabular}
\end{center}
\end{table*}

From the two right-hand panels of Figure~\ref{fig-latelc} it is
apparent that the post-explosion magnitudes of S96 are almost
identical to the pre-explosion ones in $V$, $R$, and $I$.  There is a
very slight excess in the $B$ band ($\sim 0.1$ mag), which is about
the same as the photometric uncertainty of the data.  Although it
cannot be ruled out that this excess is due to some kind of systematic
error in the calibration of the $B$-band data (the deviation from the
pre-explosion level is $\sim 1 \sigma$), it is interesting that the
ground-based $B$ and $V$ magnitudes agree very well with those
obtained by {\it Swift}/UVOT (see \S 2.2.2).

\subsubsection{Keck Spectroscopy}

\begin{figure}
\plotone{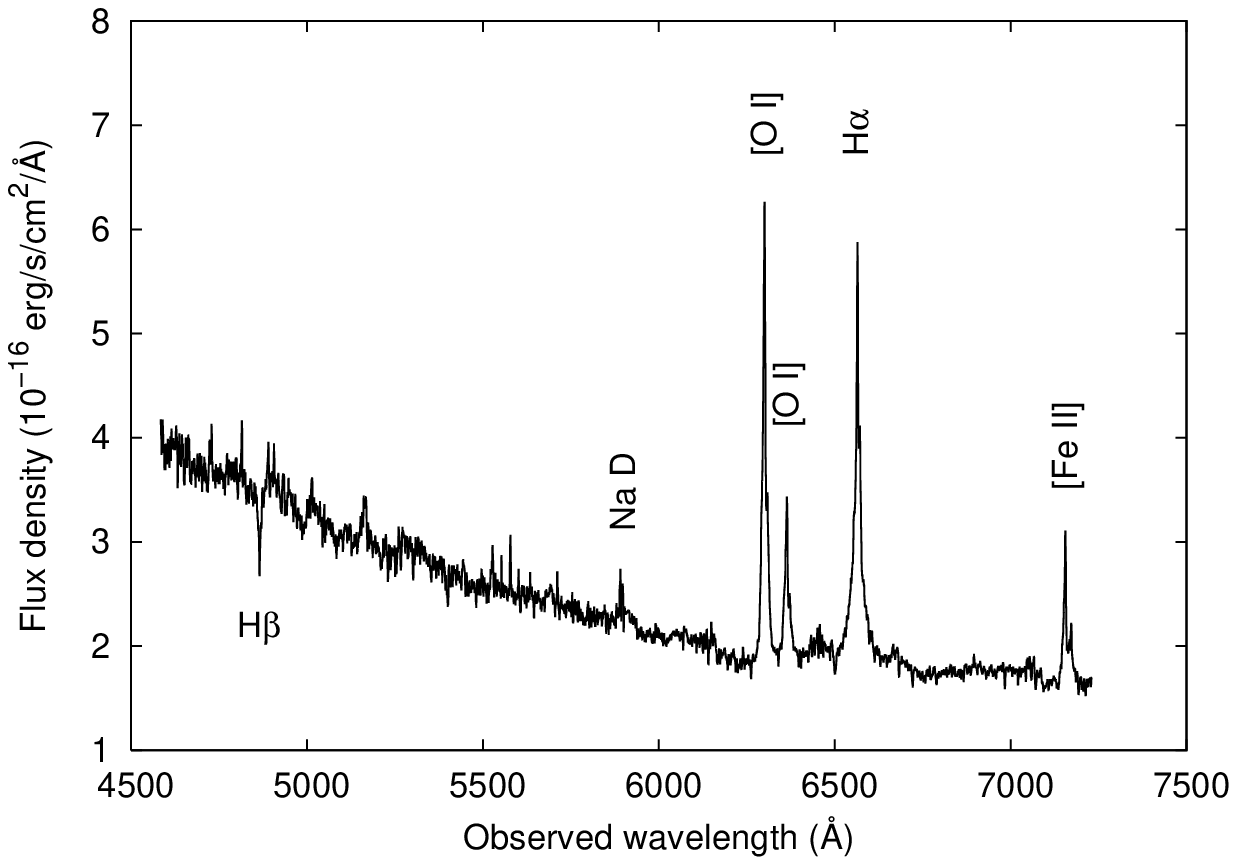}
\caption{Optical spectrum of SN~2004dj/S96 obtained with the 10~m Keck-II
telescope on 2006 Dec. 23.  The identified bright emission lines are
formed in the SN ejecta. }
\label{fig-kecksp}
\end{figure}

A late-time spectrum of SN~2004dj (exposure time 2200~s) was obtained
on 2006 Dec. 23 ($\sim 900$~d after explosion) with the DEIMOS
spectrograph \citep{Faber03} mounted on the 10~m Keck-II telescope in
Hawaii.  The 1200 line mm$^{-1}$ grating was used, with a slit $1.1''$
wide, resulting in a resolution (full width at half-maximum intensity)
of 2.7~\AA. The slit was aligned close to the parallactic angle
\citep{fil82}, so differential light losses were not a problem.

As seen in Figure~\ref{fig-kecksp}, the Keck spectrum is clearly a
composite of S96 and the nebular ejecta of SN~2004dj.  Longward of
6000~\AA, strong emission lines of H$\alpha$, [\ion{O}{1}]
$\lambda\lambda$ 6300, 6363, and [\ion{Fe}{2}] $\lambda 7155$ \AA,
characteristic of a typical nebular SN~II-P spectrum at late phases,
can be identified.  Shortward of 6000~\AA, the blue continuum
dominates the spectrum; \ion{Na}{1}~D appears in emission, which
emerges mostly from the SN ejecta, but H$\beta$ is in absorption.
Clearly, the radiation from the young stellar population of S96 is
visible in this regime.  The shape of the spectrum is fully consistent
with the predictions of population-synthesis models (see Paper~I and
\S 3).

\subsubsection{{\it HST} Observations}

\begin{figure}
\plotone{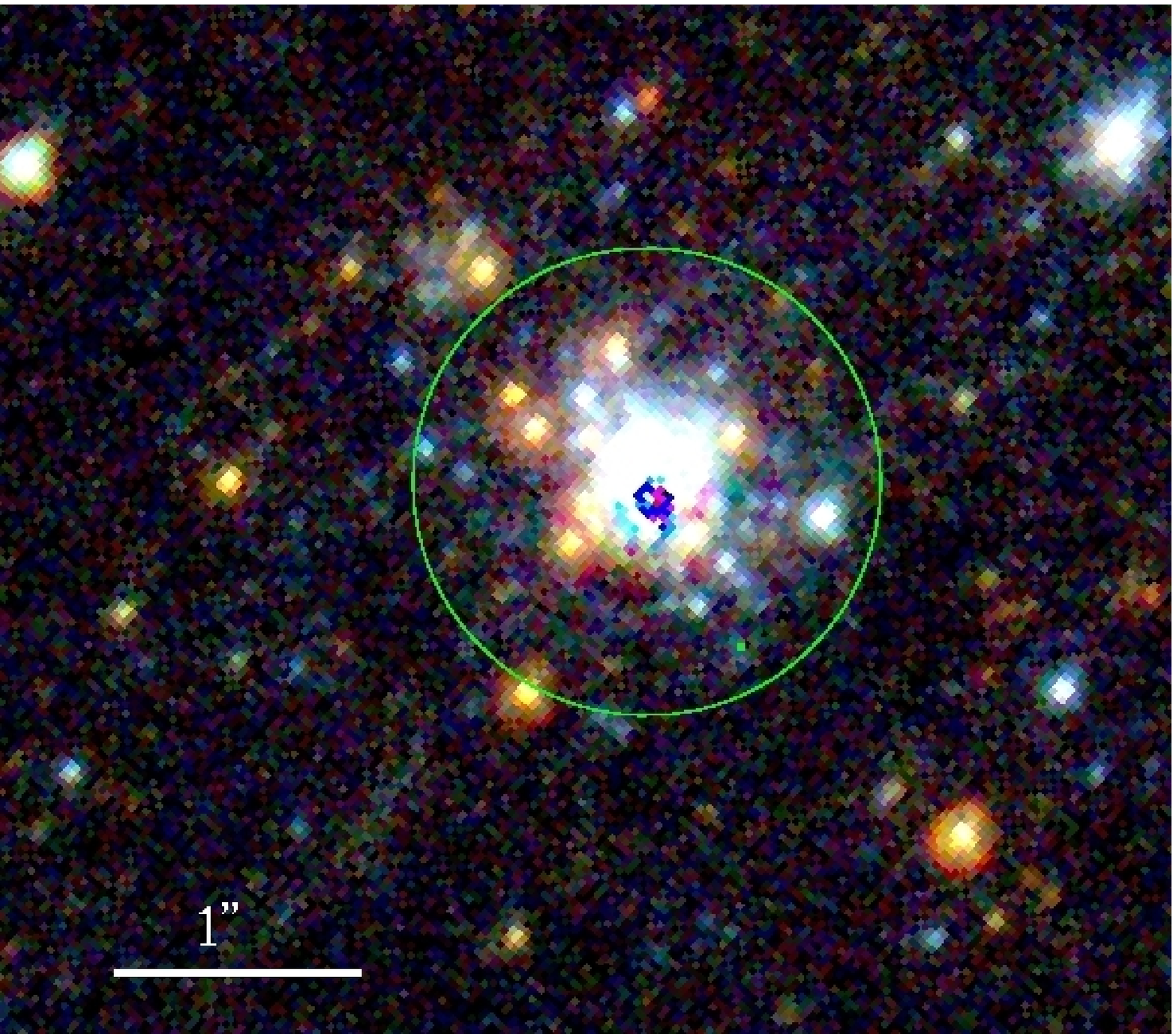}
\caption{Color-combined image from the {\it HST} $F435W$ (blue), $F606W$
(green), and $F814W$ (red) frames of S96 obtained on 2005 Aug. 28.  The
field of view is about $4'' \times 4''$; north is up, east to the
left. The green circle has a radius of 35 pixels ($\sim 15$ pc at the
distance of NGC~2403), and it defines the boundary of the cluster
region in our analysis. The PSF of SN 2004dj was modeled and removed
(see text).}
\label{fig-hst}
\end{figure}

SN~2004dj and its surrounding area was observed with {\it HST}/ACS on
2005 August 28, $\sim$425~d after explosion (GO-10607;
PI. B.~Sugerman).  Four sets of 4 drizzled frames were obtained
through the $F606W$ and $F814W$ filters, and 3 sets were recorded with
the $F435W$ filter.  In the latter case the POLUV polarization filter
was also placed in the beam.  This made it possible to study the
polarization of the SN light, but slightly complicated the photometry
of the $F435W$ frames, causing a systematic shift of the zeropoint in
the standard transformation (see \S 3.2).

The ACS frames, reduced and calibrated by the $HST$ pipeline
(including MultiDrizzle), were downloaded from the $HST$ archive at
the Canadian Astronomy Data Centre\footnote{CADC is operated by the
National Research Council of Canada with the support of the Canadian
Space Agency.}. Because SN~2004dj was still very bright compared with
the rest of S96 at the epoch of these observations, its point-spread
function (PSF) was subtracted from the ACS frames.  We used the
TinyTim
software\footnote{http://www.stsci.edu/software/tinytim/tinytim.html
.}  (version 6.3) for calculating the ACS PSFs in each filter.  Since
the analytical PSF works less effectively for drizzled frames, the
flatfield-corrected ``.FLT'' frames were used for the PSF
removal. After sub-pixel registration, the individual frames belonging
to the same filter were averaged.  The model PSF was then scaled to
the peak of the SN and subtracted from the combined frame.

The result is shown in Figure~\ref{fig-hst}. The encircled region ($r
= 35$ pixels $\approx 15$~pc) contains S96 with its unresolved inner
and resolved outer parts. Several bright red and blue giants are
visible in the outer region. The color of the unresolved inner part is
also very blue, in accord with the pre-explosion photometric
observations and the proposed young cluster age (see Paper~I). It is
also apparent that SN~2004dj occurred near the projected center of the
cluster (there are some artifacts due to the incomplete PSF removal at
the SN position, but they are less than 1\% of the subtracted SN
flux).

Photometry of the stars appearing in the ACS frames was obtained with
the DOLPHOT\footnote{http://purcell.as.arizona.edu/dolphot .} 1.0
software \citep{dolph00}. DOLPHOT incorporates corrections for
geometric distortions of the ACS camera, cosmic-ray removal, object
identification, PSF fitting (using precomputed PSFs via TinyTim),
charge-transfer efficiency correction, and transformation into
standard photometric systems. It works best with the
flatfield-corrected ``.FLT" frames. All of these frames were processed
with DOLPHOT, and the resulting magnitudes belonging to the same
filter were combined frame by frame. Only those stars that could be
identified on at least two frames with the same filter were retained
in the final list. The photometric errors were computed from the
scatter of the individual data around their mean value, taking into
account the individual magnitude errors computed by DOLPHOT. The final
magnitudes were converted to Johnson-Cousins $B$, $V$, and $I$ using
the calibration by \citet{sir05}.  Note that $0.3$ mag has been added
to the transformed $B$ magnitudes to take into account the
transmission of the $HST$ POLUV polarization filter, which was used
together with the $F435W$ filter during the observations.  The results
are analyzed in \S 3.2.

\subsection{Ultraviolet Data}

\subsubsection{{\it XMM-Newton} Observations}

\begin{table*}
\begin{center}
\caption{{\it XMM-Newton} OM Observations of S96}
\begin{tabular}{ccccccccc}
\tableline\tableline
Obs. ID & UT Date & Exp. time & $UVW2$ & Flux$^a$ &  $UVW1$ & Flux$^a$ & $U$ & Flux$^a$ \\
              &    &   (s)          & (mag)      &  &   (mag)  &  & (mag) &  \\
\tableline
0150651101 & 2003-04-30 & 6304 & 16.87 & 9.20 & 16.76 & 6.59 & 17.37 & 4.04 \\
    &        & $\sigma$  & 0.11 & 0.93 & 0.06 & 0.37 & 0.06 & 0.23 \\
\tableline
\end{tabular}
\tablecomments{$^a$The flux units are $10^{-16}$ erg s$^{-1}$ cm$^{-2}$ \AA$^{-1}$.}
\label{tbl-xmm}
\end{center}
\end{table*}

Prior to the explosion of SN~2004dj, S96 was observed with the
Optical/UV Monitor telescope (OM) on board {\it XMM-Newton}
\citep{mason01} on 2003 April 30 (PI. M.~Pakull).  The FITS frames and
tables containing the photometric data (reduced and calibrated by the
SAS pipeline) were downloaded from the XMM-Newton Science
Archive\footnote{http://xmm.esac.esa.int/xsa/index.shtml .}.  The
instrumental magnitudes of S96 (object \#1057) are listed in
Table~\ref{tbl-xmm}. Unfortunately, no $B$ or $V$ observations were
made, so full transformation into the standard Johnson system cannot
be computed. However, applying the UV transformation equations in the
OM Calibration
Documentation\footnote{http://xmm.vilspa.esa.es/external/xmm{\_}sw{\_}cal/calib/index.shtml
.}, the correction in the $U$ band is only $0.019$ mag; thus, the
instrumental magnitudes in Table~\ref{tbl-xmm} should well represent
the Vega-based standard magnitudes of S96.

Finally, the observed count rates were transformed into fluxes (in erg
s$^{-1}$ cm$^{-2}$ \AA$^{-1}$) using the conversion factors listed in
the OM Calibration Documentation (CAL-TN-0019-3-2, p.~17, Table~4). It
is known that such a conversion is only approximate, because it
depends on the SED of the object. However, the SED of S96 in the
blue/UV regime is very similar to that of an early-type star or a
white dwarf (cf. \S 3.1). The count rate to flux conversion was
calibrated using white dwarf spectrophotometric standards.  Therefore,
the flux conversion of S96 is fairly robust.

\subsubsection{{\it Swift} Observations}

\begin{table*}
\begin{center}
\caption{{\it Swift} UVOT Observations of SN~2004dj/S96}
\begin{tabular}{ccccccccccccccc}
\tableline\tableline
Obs. ID & UT Date & Exp. time & $uvw2$ & Flux$^a$ & $uvm2$ & Flux$^a$ & $uvw1$ & Flux$^a$ & $u$ & Flux$^a$ & $b$ & Flux$^a$ & $v$ & Flux$^a$\\
   &  & (s)  & (mag) &  & (mag) &  & (mag) &  & (mag) &  & (mag) &  & (mag) & \\
\tableline
00035870002 & 2006-10-09 & 2215 & -- & -- &  -- & -- & 16.96 & 6.98 & 17.44 & 3.45 & 18.02 &3.59 &17.65 & 3.24 \\
00035870003 & 2006-10-15 & 5553 & --  &  -- & --& --& 16.91 & 7.33 & 17.25 & 4.08 & 17.94 & 3.88 & 17.81 & 2.81 \\
00035870004 & 2007-04-06 & 2308 &  16.84 &9.61 & 16.91 &6.92& 17.00 & 6.78 & 17.13 &4.58 & -- & --& -- & --\\
00036563001 & 2007-12-03 & 6440 & 16.81 &9.89 & 16.80 &7.66 & 16.95 &7.09 & 17.31 &3.88 & 18.24 &2.93 & 17.76 &2.94 \\
00036563002 & 2007-12-06 & 2382 & 16.99 & 8.32 & 17.01 &6.31 & 17.07 &6.30 & 17.24 &4.13 & 18.07 &3.44& 17.80 &2.83 \\
\tableline
                        &                    & average  & 16.88 &9.27 & 16.91 &6.96 & 16.98 &6.90 & 17.27 &4.02 & 18.07 &3.46 & 17.76 &2.96 \\
   &                    & $\sigma$  &  0.10 & 0.84 & 0.11 &0.68 & 0.06 &0.39 & 0.11 &0.41 & 0.13 &0.40 & 0.07 & 0.20 \\
\tableline
\end{tabular}
\label{tbl-uvot}
\tablecomments{$^a$The flux units are $10^{-16}$ erg s$^{-1}$ cm$^{-2}$ \AA$^{-1}$.}
\end{center}
\end{table*}

The {\it Swift} Observatory \citep{geh04} was launched into orbit on
2004 November 20.  Its Ultraviolet/Optical Telescope (UVOT,
\citealt{rom05}) was used to observe SN~2004dj/S96 at five
epochs. Table~\ref{tbl-uvot} summarizes the basic parameters of these
observations. A color-combined UV image (made from the data obtained
on 2007 December 3) is presented in the right-hand panel of
Figure~\ref{fig-bok}.

The UVOT observations were downloaded from the {\it Swift} data
archive\footnote{http://heasarc.gsfc.nasa.gov/cgi-bin/W3Browse/swift.pl
.}.  The sky-subtracted frames were processed and analyzed with a
self-developed script in the following way. First, the individual
exposures belonging to the same filter (stored as extensions of the
same FITS file) were co-added with the UVOTIMSUM routine of the
HEAsoft
software\footnote{http://heasarc.gsfc.nasa.gov/docs/software/lheasoft/
.}.  For those frames that were obtained without binning, the summed
frames were rebinned by $2 \times 2$ binning ($1'' \times 1''$) in
order to enhance the signal-to-noise ratio (S/N) of the point
sources. The fluxes of S96 and the local photometric standard stars
(that could be identified on the UVOT frames) were computed with
aperture photometry in IRAF\footnote{IRAF, the Image Reduction and
Analysis Facility, is distributed by the National Optical Astronomy
Observatory, which is operated by the Association of Universities for
Research in Astronomy, Inc. (AURA) under cooperative agreement with
the National Science Foundation (NSF).}. The photometric calibration
was done according to the latest prescriptions by \citet{pool07}.  The
aperture radius was set as $5''$ (5 pixels on the $2 \times 2$ binned
frames), while the sky was computed as the ``mode" of the pixel values
in an annulus with $r_{in} = 10$ and $r_{out} = 15$ pixels centered on
the point source.  The summed, sky-corrected fluxes in
analog-to-digital units (ADU) were divided by the dead-time corrected
exposure time (defined by the keyword EXPOSURE in the FITS headers) to
obtain the raw count rates in ADU s$^{-1}$. These raw count rates were
corrected for coincidence loss following \citet{pool07}.

Finally, the corrected count rates were transformed into magnitudes
and physical fluxes by using the photometric calibrations given by
\citet{pool07}. (Note that we have applied the formulae based on the
Pickles stellar spectra, instead of gamma-ray burst spectra, because
the SED of S96 is more like that of a star than a gamma-ray burst.)
No color-term correction was applied to the magnitudes, since we
intend to compare physical fluxes rather than magnitudes from UVOT and
other instruments. The color-term corrections have been computed only
for checking the deviation of the UVOT magnitudes from the magnitudes
in the standard Johnson/Bessell system for S96 \citep{pool07}. The
corrections are $U-u = 0.22$ mag, $B-b = 0.03$ mag, and $V-v = 0.03$
mag, where lower-case letters refer to the {\it Swift} filters.  It is
seen that the UVOT $b$ and $v$ magnitudes are fairly close to the
standard system, while the $u$ magnitudes are slightly brighter.  The
final UVOT fluxes and their uncertainties are listed in Table
\ref{tbl-uvot}; they are analyzed further in \S 3.1.

\begin{figure}
\plotone{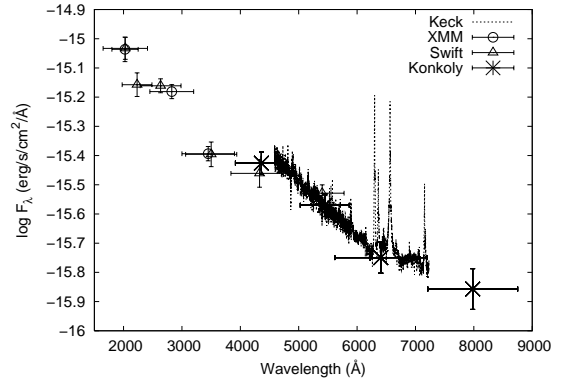}
\caption{Observed SED of S96 in the UV and optical region plotted 
against the observed wavelength. The observatories are indicated 
in the legend.}
\label{fig-uvsed}
\end{figure}

From our ground-based optical photometry (\S 2.1) it was concluded
that SN~2004dj faded below the light level of S96 by September 2006
($\sim$800~d after explosion). Because the UV flux of SNe~II-P
diminishes more rapidly than the optical flux \citep{imm07, brown07},
the UVOT observations made in October 2006 and April 2007 recorded
mostly the cluster light.  Figure~\ref{fig-uvsed} shows the comparison
of the UV fluxes with the Keck spectrum (dotted line) and ground-based
fluxes (asterisks).  The UV fluxes observed by different satellites
before and after explosion agree nicely within the errors.  The
agreement is also very good in the $B$ and $V$ bands, between the
ground-based and space-based observations; the differences are $\sim
0.06$ mag and $\sim 0.02$ mag in $B$ and $V$, respectively. Similar
agreement was obtained for some of our local photometric standard
stars, although most of them were too bright for UVOT photometry.

\section{Results}

We now present the analysis of the observations described in the
previous section.

\subsection{SED Fitting}

The physical properties of S96 were discussed by \citet{maiz04},
\citet{wang05}, and in Paper~I.  These studies were based on
pre-explosion photometry of S96 (broad-band Johnson $UBVRI$ and the
14-color BATC system in the optical, and 2MASS $JHK$ in the
NIR). Fitting the optical through NIR SED with single stellar
population (SSP) models, all of these studies revealed that S96 is a
young, compact stellar cluster with an age of $\sim$8--20 Myr.  The
uncertainty is caused by the strong age-reddening and age-metallicity
correlations in the SED fitting, and also to the sensitivity of the
stellar-evolution models applied for constructing the SED of an SSP
with a given age.

Extending the wavelength coverage of the observed SED may help break
the age-reddening-metallicity degeneracy \citep{rebuz86, kav07}.  The
SED of young stellar clusters can be best characterized in the UV,
because the UV luminosity, originating from the most massive,
fast-evolving supergiants, strongly correlates with the cluster age
\citep{ocon99, buz07}. By adding the UV data from {\it XMM-Newton} and
{\it Swift} (\S 2.2) to the optical through NIR SED used in the
previous studies, one can get a better constraint for the cluster age,
and hence the SN mass.

There is, however, an additional, non-negligible source of systematic
uncertainty in the interpretation of the SEDs of young massive
clusters: they contain $\sim 10^4$--$10^5$ stars, and the high-mass end
of their initial mass function (IMF) is poorly populated.  Because
these stars are also the most luminous ones, the observed SEDs of
clusters having the same physical parameters will statistically
deviate from the model SED depending on the actual number of their
most massive, luminous stars \citep{cervi02, cervi04, jamet04,
cervi06}.  This clearly limits the applicability of model SEDs
computed by assuming an analytical, completely sampled IMF (i.e.,
practically an infinite number of stars). Because this effect was not
taken into account in the previous studies of S96, here we address it
in detail.

In Table~\ref{tbl-sed} we have collected all available photometric
data (converted to fluxes in erg s$^{-1}$ cm$^{-2}$ \AA$^{-1}$) of
S96, including both pre-explosion and post-explosion observations.  We
have used our $BVRI$ photometry made $800$~d after explosion
(Table~\ref{tbl-phot}) for the representation of the post-explosion
flux in the optical. Unfortunately, there are no post-explosion
observations in the $JHK$ bands at our disposal. 

\begin{table}
\footnotesize
\begin{center}
\caption{Time-Averaged SED of S96 Before and After SN~2004dj. }
\label{tbl-sed}
\begin{tabular}{ccccl}
\tableline\tableline
Filter & $\lambda_c$ & $\Delta \lambda$ & $F_\lambda$ & References \\
\tableline
\multicolumn{5}{c}{Before explosion} \\
\tableline
$UVW2$ & 2025 & 450 & 9.20 (0.93) & This paper ({\it XMM})\\
$UVW1$ & 2825 & 750 & 6.59 (0.37) & This paper ({\it XMM})\\
a & 3360 & 360 & 3.11 (0.62) & \cite{wang05} \\
$U_{OM}$ & 3450 & 900 & 4.04 (0.23) & This paper ({\it XMM})\\ 
$U$ & 3663 & 650 & 3.58  (0.43) & \cite{larsen99} \\
c & 4210 & 320 & 3.28 (0.12) & \cite{wang05} \\
$B$ & 4361 & 890  & 3.33 (0.29) & Paper~I \\
d & 4540 & 340 & 2.88 (0.08) & \cite{wang05} \\
$g'$ & 4872 & 1280 & 2.74 (0.10) & \cite{dav07} \\
e & 4925 & 390 & 2.54 (0.16) & \cite{wang05} \\
f & 5270 & 340 & 2.22 (0.10) & \cite{wang05} \\
$V$ & 5448 & 840  & 2.72  (0.11) & Paper~I \\
g & 5795 & 310 & 2.18 (0.08) & \cite{wang05} \\
h & 6075 & 310 & 2.14 (0.11) & \cite{wang05} \\
$r'$ &  6282 & 1150 & 1.93 (0.10) & \cite{dav07} \\
$R$ & 6407 & 1580 & 1.75 (0.15) & Paper~I \\
i & 6656 & 480 & 1.90 (0.10) & \cite{wang05} \\
j & 7057 & 300 & 1.98 (0.16) & \cite{wang05} \\
k & 7546 & 330 & 1.79 (0.09) & \cite{wang05} \\
$i'$ &  7776 & 1230 & 1.64 (0.10) & \cite{dav07}\\
$I$ &  7980 & 1540 & 1.37 (0.17) & Paper~I \\
m & 8023 & 260 & 1.59 (0.11) & \cite{wang05} \\
n & 8480 & 180 & 1.50 (0.23) & \cite{wang05} \\
o & 9182 & 260 & 1.35 (0.13) & \cite{wang05} \\
p & 9739 & 270 & 1.24 (0.25) & \cite{wang05} \\
$J$ &12200 & 2130 & 1.06 (0.08) & \cite{2mass}\\
$H$ &16300 & 3070 & 0.70 (0.07) & \cite{2mass}\\
$K$ &21900 & 3900 & 0.28 (0.05) & \cite{2mass}\\
\tableline
\multicolumn{5}{c}{After explosion} \\
\tableline
$uvw2$ & 2030 & 760 & 9.27 (0.84) & This paper ({\it Swift})\\
$uvm2$ & 2231 & 510 &  6.96 ( 0.68) & This paper ({\it Swift})\\
$uvw1$ & 2634 & 700 & 6.90 (0.39) & This paper ({\it Swift})\\
$u$ & 3501 & 875 & 4.02 (0.41) & This paper ({\it Swift})\\
$b$ & 4329 & 980 & 3.46 (0.40) & This paper ({\it Swift})\\
$v$ & 5402 & 750 & 2.96 (0.20) & This paper ({\it Swift})\\
$B$ & 4361 & 890 & 3.75 (0.34) & This paper \\
$V$ & 5448 & 840 & 2.69 (0.25) & This paper \\
$R$ & 6407 &1580 & 1.78 (0.23) & This paper \\
$I$ &  7980 & 1540 & 1.39 (0.24) & This paper \\
\tableline
\end{tabular}
\tablecomments{$\lambda_c$ and $\Delta \lambda$ denote the central
wavelength and the full width at half-maximum intensity (FWHM) of a 
given filter, in \AA. The flux units are
$10^{-16}$ erg s$^{-1}$ cm$^{-2}$ \AA$^{-1}$. Uncertainties are given 
in parentheses.}
\end{center}
\end{table}

\begin{figure}
\plotone{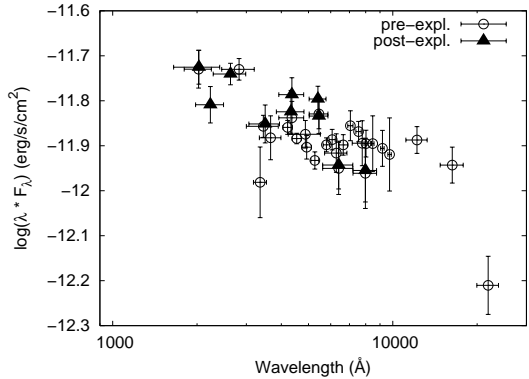}
\caption{Comparison of pre- and post-explosion SEDs of S96.}
\label{fig-prepost}
\end{figure}

The pre- and post-explosion SEDs are plotted together in
Figure~\ref{fig-prepost}.  It is apparent that the two datasets agree
within the errors. The agreement in the UV and optical bands implies
that the removal of the flux of the progenitor of SN~2004dj from the
integrated light of the cluster has not caused a significant loss of
light in these bands.

In order to fit theoretical SEDs to the observations, we have defined
an averaged ``normal'' SED of S96 by combining the pre- and
post-explosion data. In the UV range, between $2000$ and $4000$~\AA,
we adopted the average of the fluxes from {\it XMM-Newton}/OM and {\it
Swift}/UVOT. The ground-based data in this spectral range are
expected to be less reliable than the satellite-based ones, because of
the higher probability of systematic errors introduced by the local
atmospheric conditions.  In the optical we used the Johnson-Cousins
$BVRI$ data. In the NIR, only the pre-explosion 2MASS $JHK$ fluxes
\citep{2mass} were available to us. 

There is a possibility that the observed SED is somewhat contaminated
by foreground/background stars belonging to NGC~2403, altering the
fluxes from being entirely due to a SSP. The amount
of this contamination is difficult to estimate, because S96 itself
may contain some older stars captured from its
galactic neighborhood (see \S 3.2 and \S 4). However, the background
subtraction we applied during photometry should have removed most of
the flux from background stars. Due to the compactness of
S96 ($\lesssim 30$ pc diameter, Fig.\ref{fig-hst}), the
number of foreground field stars should be minimal 
(see \S 3.2 for a detailed discussion). Consequently, the SED fluxes
are expected to be due mostly to S96. Since S96 is by far the brightest
source in this region, the contamination from field stars should
not exceed the estimated errors of the ``normal'' SED 
fluxes, $\sim 10$\%.

During the analysis the distance of NGC~2403 was fixed at 3.5~Mpc.  
This optimal value is found by combining various 
distance measurement results for SN~2004dj and its host galaxy, 
as discussed in Paper~I.

In order to test whether the effect of statistical IMF sampling allows
the modeling of the cluster SED, \citet{cervi04} introduced the
concept of the ``lowest luminosity limit'' (LLL).  It can be simply
expressed as follows: the integrated luminosity of the cluster must be
higher than that of the most luminous star of the model isochrone, at any
wavelengths.  The LLL is a strong function of age and it also depends
on the considered wavelength regime (or filter band). As expected, LLL
gives the strongest constraint for the models with age 1--100~Myr
\citep{cervi04}. Because the possible ages of S96 found in
previous studies are all in this interval, it is necessary to check
whether the cluster meets this criterion. The normal SED fluxes
described above were dereddened and converted to absolute magnitudes
assuming a distance of 3.5~Mpc and $E(B-V) = 0.07$ mag (see
Paper~I). This resulted in $M_B \approx -9.2$ mag, $M_R \approx -10.0$ mag
and increasing up to $M_K \approx -12.3$ mag. The predictions for the
most massive stars from the Padova isochrones \citep{cio06a, cio06b}
with $t = 8$ Myr (the youngest age proposed for S96 so far) are
$-7.3$, $-9.1$, and $-11.0$ mag ($Z=0.008$) and $-8.2$, $-9.1$, and
$-10.1$ mag ($Z=0.019$) for the $B$, $R$, and $K$ bands,
respectively. We see that S96 is at least $\sim 1$ mag brighter in
the optical/NIR bands, and it is also $\sim 2$ mag brighter in
$U$. This criterion becomes more relaxed toward higher ages as the
possible most luminous stars become fainter. Our first conclusion is
that although S96 is close to the LLL for 8 Myr, it is definitely
above it, so this cluster is sufficiently rich to make the comparison of
its observed SED with theoretical models statistically feasible.

For comparison with the observations, 
we have applied three different classes of SSP models with somewhat different input physics.
First, as in Paper~I, we selected the GALAXEV models of \citet{bruz03}
that are based on the Padova evolutionary tracks.  Second, we chose
the SPEED models by \citet{jim04} that were computed using a new set
of stellar interior models, evolutionary tracks, and different
treatment of the mass loss. For both of these models, a Salpeter
IMF Was adopted, similarly to Paper~I and
previous studies.  The machine-readable data of these two model sets
were downloaded from the SPEED
website\footnote{http://www.astro.princeton.edu/$\sim$raulj/SPEED/index.html 
.}. Third, we applied the SSP models generated by the
Starburst99\footnote{http://www.stsci.edu/science/starburst99/ .}
code \citep{vaz05}. The Starburst99 models are highly configurable;
the user may choose between different evolutionary tracks, atmospheric
models, and pre-computed spectral libraries to create a unique set of
SSP models.  In order to test the model dependency of the results, we
have chosen the Geneva evolutionary tracks and Kroupa IMF, and generated
SSP SEDs using metallicities $Z=0.004$, $0.008$, and $0.02$, between
$t=0$ and $100$ Myr. Note that the metallicity resolution of the SPEED
models is lower; only models with $Z=0.004$ and $0.02$ are
available. The age step of the Starburst99 models was selected as
$\Delta t = 1$ Myr sampling linearly between 1 and 100 Myr, providing 
much better age resolution than the SPEED models for $t > 10$ Myr.

The model SEDs were compared with the observations via the usual 
$\chi^2$ fitting. Note that throughout the paper we use the
reduced $\chi^2$ (the sum of the squares of residuals divided
by the number of data points). The optimized parameters were 
the cluster mass $M_c$, the cluster age $T_c$, and the
reddening $E(B-V)$, while the metallicity and the distance 
were kept fixed. The interstellar extinction
at any wavelength in the considered UVOIR spectral regime was
calculated adopting $R_V = 3.1$ and the average Galactic reddening 
law of \citet{fitz07}. 

First, the fitting was computed in the canonical way, not taking into
account the statistical fluctuations in the IMF sampling.
and the models were fitted directly to the observed normal SED, as if the
cluster were composed of an infinite number of stars. In this
case the $\chi^2$ function was defined as 
\begin{equation}
\chi^2 ~=~ {1 \over N_{obs}} \sum_{i=1}^{N_{obs}} {1 \over \sigma_i^2}
[ F^{obs}(\lambda_i) - M_c \cdot S^{mod}(\lambda_i, T_c) ]^2,
\end{equation}
where $N_{obs}$ is the number of observed points in the normal SED,
$F^{obs}$ is the observed flux at wavelength $\lambda_i$ (corrected
for distance and extinction), $\sigma_i$ is its uncertainty, and
$S^{mod} (\lambda_i, T_c) $ is the flux of the model SED with age
$T_c$, at the same wavelength.  Because the model fluxes are usually
normalized to 1~M$_{\odot}$, the cluster mass $M_c$ enters simply as a
scale parameter in this expression.  The results of these calculations
are summarized in Table \ref{tbl-sedfit1}.  The age resolution for the
SPEED models was slightly increased by interpolating between the two
neighboring model SEDs for those ages that were not covered by the
original models. However, without this correction, the best-fitting
parameters did not change significantly. The cluster parameters
inferred in this way are very similar to the results of earlier
investigations cited above.

\begin{table}
\begin{center}
\caption{Parameters of the SED Fitting, Without IMF Fluctuations (See text)}
\label{tbl-sedfit1}
\begin{tabular}{lccccc}
\tableline\tableline
Model & $Z$ & $T_c$ & $M_c$ & $E(B-V)$ & $\chi^2$ \\
  &  &  ($10^6$ yr) & ($10^3 ~M_\odot$) & (mag) & \\
\tableline
J04 & 0.004 & 8 & 27 & 0.09 & 1.422 \\
J04 & 0.020 & 24 & 99 & 0.04 & 2.313 \\
BC03 & 0.004 & 35 & 114 & 0.13 & 2.794 \\
BC03 & 0.020 & 26 & 92 & 0.13 & 0.536 \\
BC03 & 0.020 & 9 & 37 & 0.17 & 0.886 \\
SB99 & 0.004 & 40 & 90 & 0.08 & 3.279 \\
SB99 & 0.008 & 9 & 26 & 0.12 & 3.521 \\
SB99 & 0.020 & 9 & 24 & 0.10 & 1.624 \\
SB99 & 0.020 & 40 & 91 & 0.07 & 3.918 \\
\tableline
\end{tabular}
\end{center}
\end{table}

Second, the statistical IMF sampling was taken into account as
follows.  For each age and wavelength, an uncertainty
$\sigma_{mod}(\lambda_i, T_c)$ was assigned for any model flux as a
measure of the fluctuation of the model SED fluxes due to the random
sampling of the IMF. Then, each model flux was modified as
$S(\lambda_i) \pm \xi$ where $0 < \xi < \sigma_{mod}(\lambda_i,T_c)$
is a random variable. This step was repeated $N_{mod}$ ($=1000$)
times, thus constructing a series of model SEDs that fluctuate around
the original model fluxes. For the $k$th model ($1 \leq k \leq
N_{mod}$), $\chi_k^2$ was computed as in Eq. (1). 
Finally, following
the recommendation by an anonymous referee, the final $\chi^2$ was
determined as 
\begin{equation} \chi^2 ~=~ - {2 \over N_{obs}} \ln
\left ( {1 \over N_{mod}} \sum_{k=1}^{N_{mod}} P_k \right ),
\end{equation} 
where $P_k = \exp(-0.5 N_{obs} \chi_k^2)$ is the
likelihood that the $k$th model describes the observations. Eq. (2) means
that the $P_k$ likelihoods are averaged, and the final $\chi^2$ is
computed from $P_{ave}$. This approach gives lower final $\chi^2$ 
values than the simple average of the individual $\chi_k^2$ values.

The modification of the $\chi^2$ function in Eq. (2) ensures that the
$\chi^2$ is mostly sensitive to those models that are particularly
affected by the sampling effect (i.e., those whose $\sigma_{mod}$ is
high) but give a good fit to the observations, while giving lower
weight to those models that produce inferior fits.  Also, the $\chi^2$
value remains mostly unchanged when the random sampling effect is
negligible, because in this case the individual $\chi_k^2$ values (and
the corresponding $P_k$s) are nearly the same for each random model.
 
Of course, the reliability of this approach depends heavily on the
proper selection of the $\sigma_{mod}$ values. Moreover, the
$\sigma_{mod}$ values belonging to different filters are correlated,
because the addition or subtraction of one bright star would affect
the cluster flux in all bands. This correlation is not reflected by
our random models, as the fluctuations were added to the fluxes
independently in each band. The proper treatment of all these effects
would require full computation of detailed SSP models, taking into
account the statistical population of the IMF, which is, however,
beyond the scope of this paper.

Instead, we followed a simpler approach as a first approximation: a
Salpeter IMF between 0.5 and 100~M$_\odot$ was randomly populated
with stars until the sum of their mass reached $M_c = 10^5$ M$_\odot$
(roughly the mass of S96). The luminosities of these stars in the
$UBVRIJHK$ bands were selected from Padova isochrones at a given age
(between 4 and 100 Myr) and simply summed up to get an estimate of
the cluster SED. One thousand such SEDs with the same cluster mass and age
were generated and the standard deviations of the fluxes were computed
at each wavelength.

\begin{figure}
\plotone{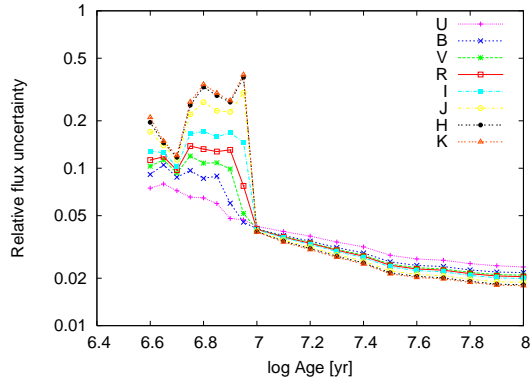}
\caption{The relative SED flux uncertainty due to IMF fluctuations as a function of
age. Different symbols are used for different filter bands, as indicated by the labels in
the upper-right corner.}
\label{fig-stat}
\end{figure}

In Figure~\ref{fig-stat} the relative flux uncertainty (defined as
$\sigma_{mod} / S$, where $S$ is the mean flux of the SED at a given
wavelength and $\sigma_{mod}$ is its standard deviation) for each band
is plotted as a function of age for $Z=0.019$. The results are very
similar in the case of $Z=0.008$ and $Z=0.004$.  It is apparent that
the relative uncertainty drops below 5\% for $t > 10$ Myr, but it may
reach up to 30\% for younger clusters at certain bands. Generally,
the $U$ and $B$ bands are less affected (the maximum uncertainty is
$<$10\%), but the scatter becomes higher toward
the NIR. This means that the SED of such young clusters would
fluctuate from cluster to cluster with amplitudes up to 30\% of the
flux values.  The reason for the drop of the fluctuations above $10$
Myr is that the IMF fluctuations are most pronounced at 
$M \geq 20$ M$_\odot$ for a $10^5$ M$_\odot$ cluster considered here, and
the highest possible mass decreases quickly below this limit for 
$t > 10$ Myr.

\begin{table*}
\begin{center}
\caption{Relative Flux Uncertainties $\sigma_{mod}/S_{mod}(\lambda_i,Tc)$.$^a$}
\label{tbl-sigma}
\begin{tabular}{cccccccccccc}
\tableline\tableline
Age & $uvw1$ & $uvm2$ & $uvw1$ & $U$ & $B$ & $V$ & $R$ & $I$ & $J$ & $H$ & $K$ \\
$< 10$ Myr & 0.05 & 0.05 & 0.05 & 0.08 & 0.10 & 0.10 & 0.15 & 0.20 & 0.25 & 0.30 & 0.30 \\
$> 10$ Myr & 0.03 & 0.03 & 0.03 & 0.03 & 0.03 & 0.03 & 0.03 & 0.03 & 0.03 & 0.03 & 0.30 \\
\tableline
\end{tabular}
\tablecomments{$^a$The values for the UV bands are based on
extrapolation, since these bands were not covered by the applied
isochrones.}
\end{center}
\end{table*}


The relative flux uncertainties 
($\sigma_{mod}/S(\lambda_i)$) are listed in 
Table \ref{tbl-sigma}. These values were applied in
the computations of the modified $\chi^2$, as discussed above.
The results of the $\chi^2$ minimizations are in Table \ref{tbl-sedfit2}.
Note that in this case the interpolations between ages were not applied at all, 
and the fitting was computed only for those ages that were covered by 
the original models. 
However, when we used the simple average of the $\chi_k^2$ 
values as the final $\chi^2$, instead of the $P_k$ likelihoods as in Eq. (2),
the parameters of the best-fitting models did not change. 

\begin{table}
\begin{center}
\caption{Parameters of the SED Fitting}
\label{tbl-sedfit2}
\begin{tabular}{lccccc}
\tableline\tableline
Model & $Z$ & $T_c$ & $M_c$ & $E(B-V)$ & $\chi^2$ \\
  &  &  ($10^6$ yr) & ($10^3$ M$_\odot$) & (mag) & \\
\tableline
J04 & 0.004 & 8 & 34 & 0.11 & 1.802 \\
J04 & 0.02 & 20 & 94 & 0.04 & 2.511 \\
BC03 & 0.004 & 40 & 121 & 0.10 & 2.973 \\
BC03 & 0.02 & 20 & 86 & 0.14 & 0.629 \\
BC03 & 0.02 & 8 & 34 & 0.18 & 1.453 \\
BC03 & 0.02 & 10 & 41 & 0.15 & 1.492 \\
SB99 & 0.004 & 30 & 72 & 0.09 & 3.257 \\
SB99 & 0.008 & 14 & 30 & 0.08 & 3.018 \\
SB99 & 0.02 & 14 & 39 & 0.11 & 1.497 \\
SB99 & 0.02 & 28 & 72 & 0.09 & 2.666\\
\tableline
\end{tabular}
\tablecomments{$^a$IMF fluctuations taken into account (see text).}
\end{center}
\end{table}

\begin{figure*}
\plottwo{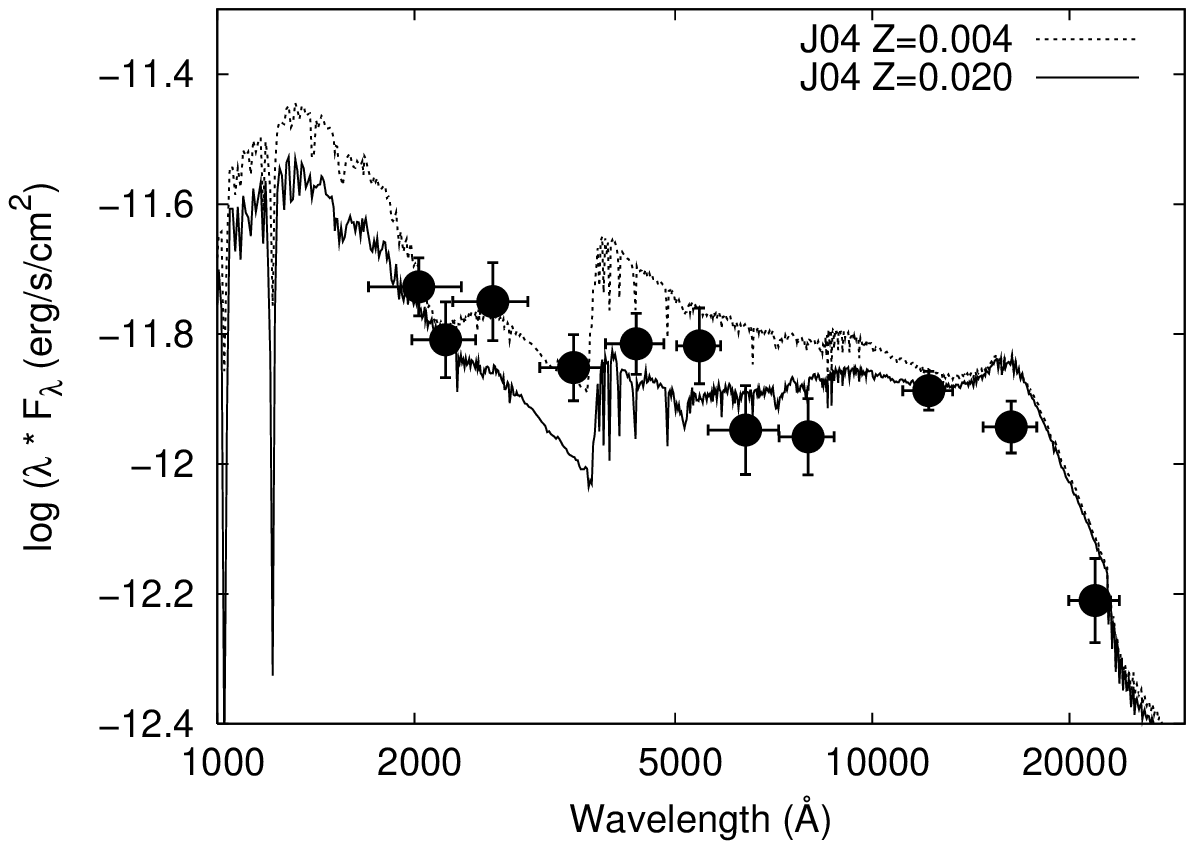}{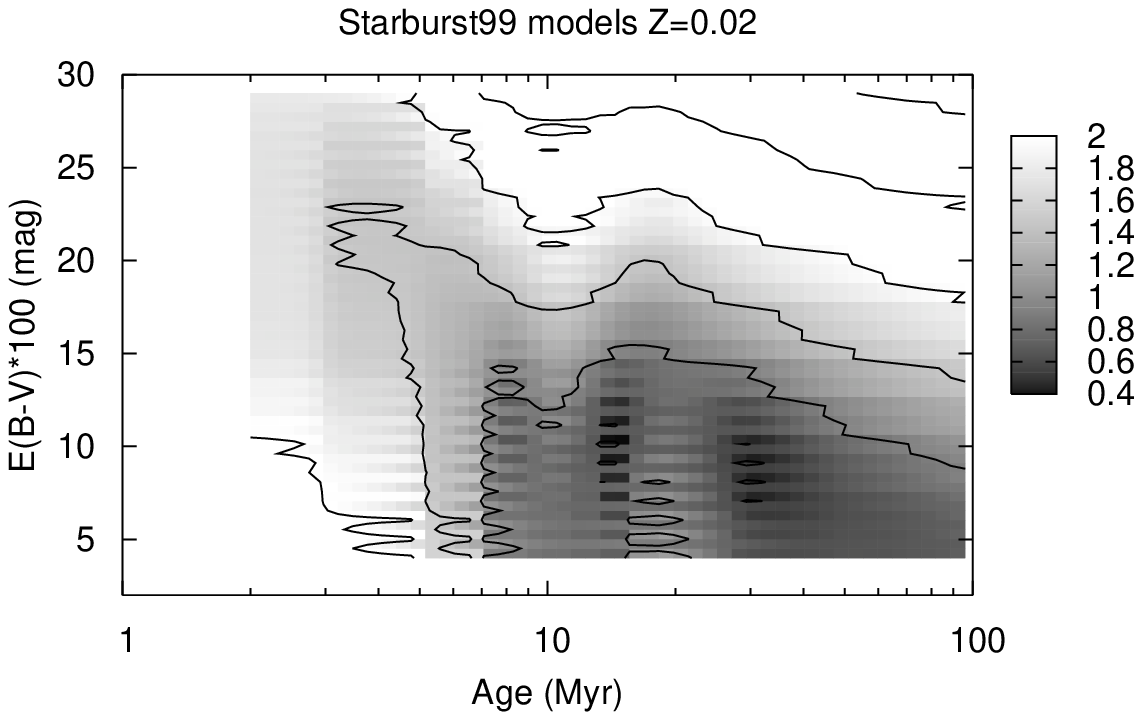}
\plottwo{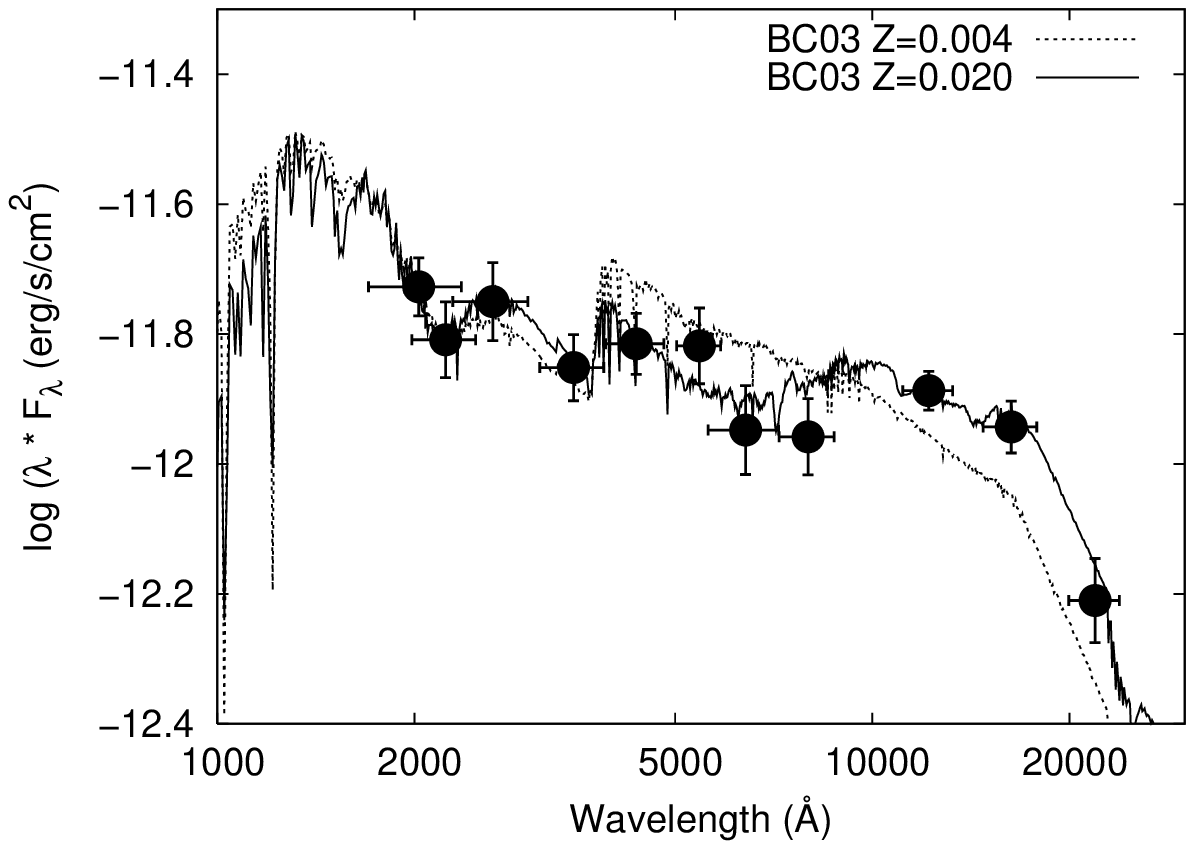}{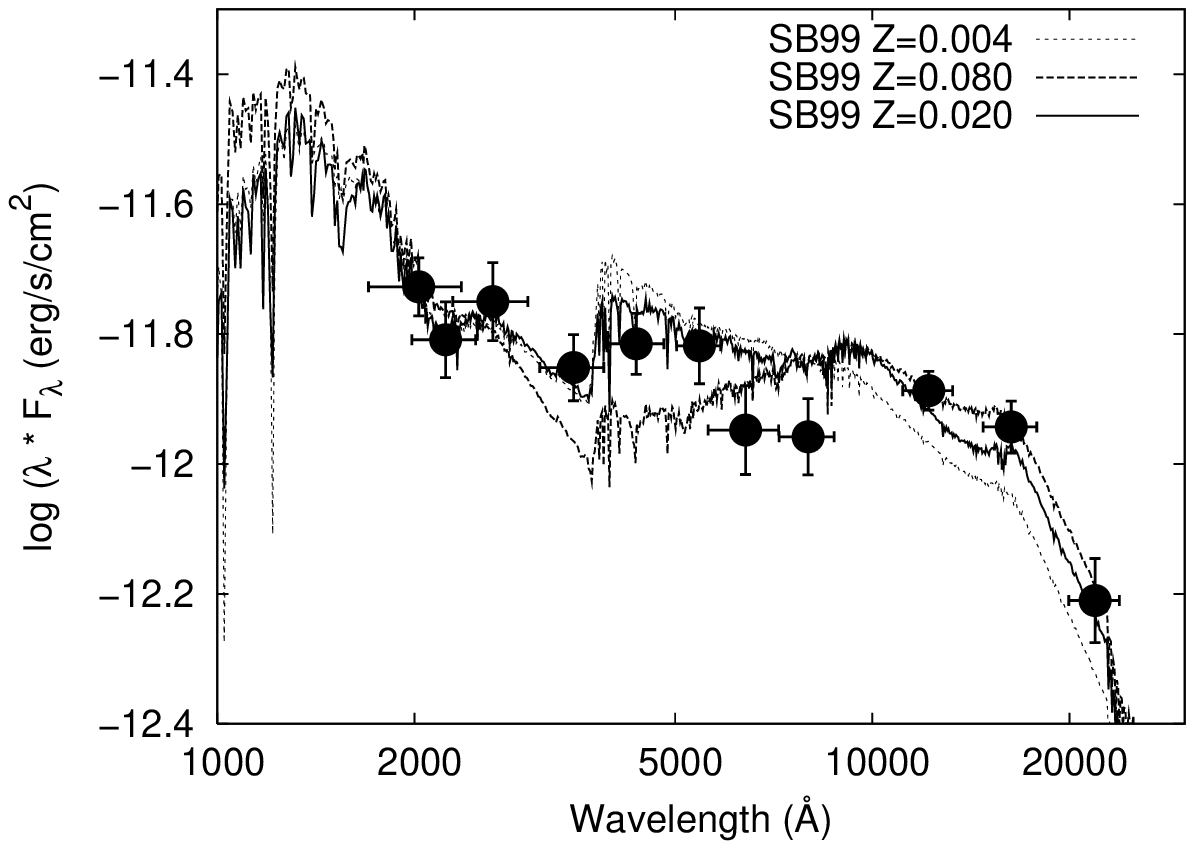}
\caption{SEDs of SSP models fitted to the observations of S96 with IMF sampling 
effects taken into account.  Filled circles represent the ``normal'' SED of the cluster 
constructed by averaging the observations within the selected wavelength bands (see
text).  The horizontal error bars denote the wavelength range of the
broad-band filters, while the vertical error bars correspond to the
uncertainties of the data.  Solid and dotted lines show the mean fluxes of the
theoretical SEDs with parameters indicated as labels.  The top-right
panel presents the grayscale map of a $\chi^2$ space of a bimodal solution, 
where the dark areas represent the most probable solutions (the horizontal stripes
are due to the finite resolution of the grid).}
\label{fig-sedfit}
\end{figure*}

Figure~\ref{fig-sedfit} shows the results of the SED fitting with the
IMF sampling effects taken into account (note that only the mean
fluxes of the best-fitting models SEDs are plotted). The fitting of
the unperturbed SEDs (i.e., ignoring the IMF fluctuations) resulted in
very similar figures.  In general, the models applied in this study
give an adequate representation of the observed SED of S96 with the
cluster parameters collected in Tables~\ref{tbl-sedfit1} and
\ref{tbl-sedfit2}.

In many cases, the $\chi^2$ map (upper-right panel of
Fig.~\ref{fig-sedfit}) showed not one, but two 
distinct minima at two different ages, regardless of
the presence or absence of IMF fluctuations. 
This was first noted by \citet{maiz04}, and it is  
confirmed here. \citet{maiz04}
found that their younger solution ($T_c \approx 14$ Myr) 
had the lower $\chi^2$ of
the two minima. In the present case, it turned out to be 
model dependent.  In the case of the BC03 models, 
the older solution has slightly lower $\chi^2$, 
while for the SB99 models it is the younger one that
has a deeper minimum. 

\begin{figure}
\plotone{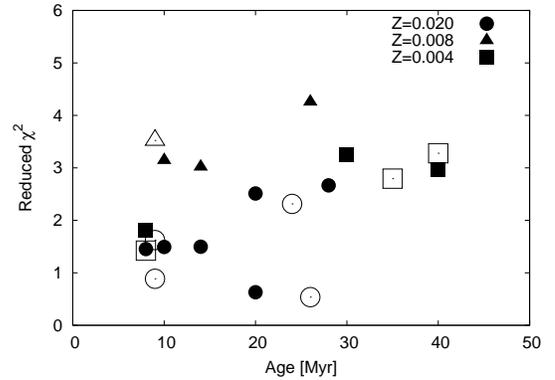}
\caption{The $\chi^2$ of the best-fitting model SEDs as a function of
cluster age (in Myr).  Filled symbols denote the models with IMF
fluctuations (Table~\ref{tbl-sedfit2}), while open ones indicate the
canonical solutions (without random IMF population,
Table~\ref{tbl-sedfit1}). Different symbol shapes denote different
metallicities as indicated in the upper-right corner.}
\label{fig-sedres}
\end{figure}

Figure~\ref{fig-sedres} shows the $\chi^2$ of the best-fitting models
(i.e., those listed in Tables~\ref{tbl-sedfit1} and \ref{tbl-sedfit2})
plotted as a function of the cluster age $T_c$.  It is apparent that
all the models with the lowest $\chi^2$ have solar metallicity
($Z=0.02$) except one model by J04 which has $Z=0.004$.  Also, the
canonical model fits (without IMF fluctuations, open symbols) clearly
show the two-age structure noted previously: a ``young'' solution with
$T_c \approx 8$--10 Myr and an ``older'' one with $T_c \approx 25$--40
Myr, preferring the ``young'' solutions on the basis of $\chi^2$.  On
the other hand, the models with IMF fluctuations (filled symbols)
present a continuous distribution between the same age limits, also
with a preference for the younger, solar-metallicity models.  The
disappearance of the bimodal distribution is mainly due to the fact
that when random IMF sampling is taken into account, the younger, less
massive models are more affected, which may increase their
$\chi^2$. (Though there might be a few models that fit the
observations very well, this is probably not true for the majority of
them.) This gives higher preference to the models that are less
affected by random sampling, thus shifting the best-fitting models
toward those with $T_c \geq 10$ Myr.

Although the lowest-$\chi^2$ models have $Z=0.02$, this may be
misleading, because the metallicity is a weakly constrained parameter
in SED fitting.  Additional information about the possible cluster
metallicity may help strengthen the confidence of this
parameter. Independent observations suggest that the average
metallicity of NGC~2403 is below solar.  The oxygen abundance in
NGC~2403 at the position of S96 is [O/H] = $-0.24$, from spectroscopy
of \ion{H}{2} regions \citep{pil04}.  The distribution of red
supergiants in the color-magnitude diagram of stars within the inner
disk \citep{dav07} also suggests that the average metallicity of the
whole population is $Z \approx 0.008$. On the other hand,
\citet{maiz04} preferred solar metallicity for S96 based on its
galactocentric distance and the abundance gradient in NGC~2403 found
by \citet{fierro86}.  There seems to be no clear consensus on the
possible metallicity of S96. The fitting of the cluster SED as a whole
suggests $Z \approx 0.02$, but with high uncertainty.  This problem is
investigated further in the next section with the analysis of the
resolved stellar population of S96.

Contrary to previous solutions (Paper~I), the reddening parameter is
much more tightly constrained in this case. Its average value is
$E(B-V) \approx 0.10 \pm 0.05$ mag, while in previous studies values
as high as $E(B-V) \approx 0.35$ mag were also proposed. Also, the
age-reddening-metallicity degeneracy is much reduced in this case,
owing to the increased wavelength coverage of the observed SED in the
UV. This reddening value is in very good agreement with
$E(B-V)_{SN}=0.07 \pm 0.1$ mag derived for SN~2004dj
(Paper~I). Although the SED of the whole cluster can, in principle, be
much more affected by intracluster reddening than SN~2004dj itself if
the position of the SN within the cluster is on the near side toward
the observer, our new $HST$ observations also strongly suggest that
$E(B-V) \approx 0.1$ mag for all the stars resolved within the cluster
(\S 3.2). We adopt $E(B-V) = 0.1 \pm 0.05$ mag for the rest of this
paper.

Looking for additional constraints on the cluster parameters, we
attempted to fit the high-resolution Keck spectrum (\S 2.1.2) with
high-resolution SSP model spectra by \citet{del05} that are based on
the Geneva tracks.  The motivation for this was the sensitivity of
some spectral features on the cluster age and metallicity
\citep{koleva08}. However, this analysis was complicated by the obvious
presence of the SN nebular lines that dominate the red part of the
integrated spectrum. Because H$\beta$ is in absorption, its origin
should be mostly from cluster stars, but the contribution from
SN~2004dj may be non-negligible. It is difficult to estimate the SN
contamination at $\sim 900$~d, because very few observed SN spectra
exist at this epoch. We examined three spectra of SN~1987A that were
taken 700--1000~d past explosion \citep{pun95}, and found that there
is an emission feature that may be attributed to H$\beta$ in these
spectra. The amplitude ratio of these emission components, 
H$\alpha$/H$\beta$, is found to be $\sim 4$. 

The SN~2004dj contamination at H$\beta$ in the Keck spectrum was estimated
in the following way. First, a Lorentzian emission profile was fitted to the observed 
H$\alpha$ line, taking into account the absorption component of the
other cluster stars from the SSP models (note that the H$\alpha$ emission 
due to SN~2004dj is so strong that the absorption component has only a minor
effect on the fitted amplitude). Second, this profile was shifted to
the rest wavelength of H$\beta$ with its amplitude divided by the
H$\alpha$/H$\beta$ amplitude ratio (AR, using AR$=4$ as default) and its
damping parameter $\gamma$ multiplied by $0.548$ (taking into account 
that the Lorentzian FWHM for pressure broadening scales with $\lambda^2$). 
The fitting was recomputed using different amplitude
ratios between $3$ and $6$, to test the sensitivity of the results on
this parameter.

\begin{figure*}
\plottwo{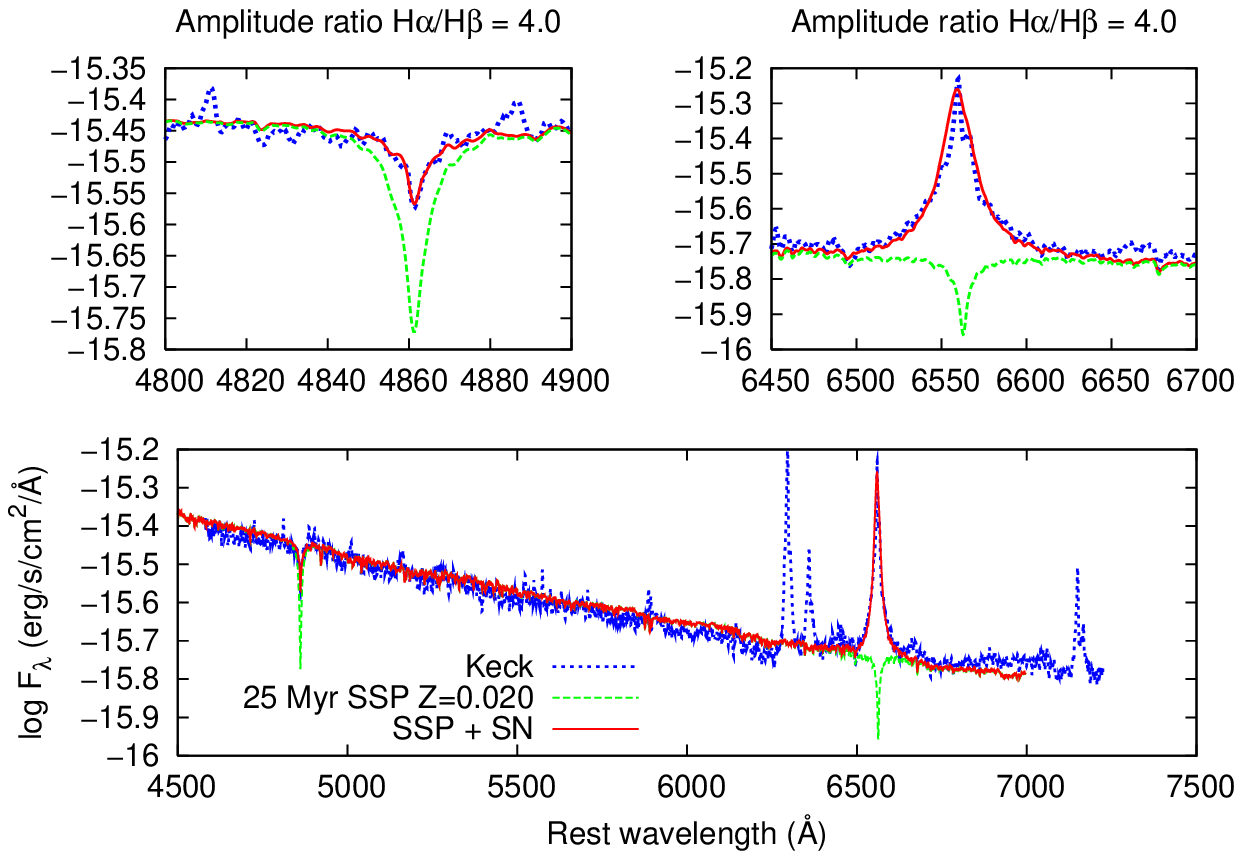}{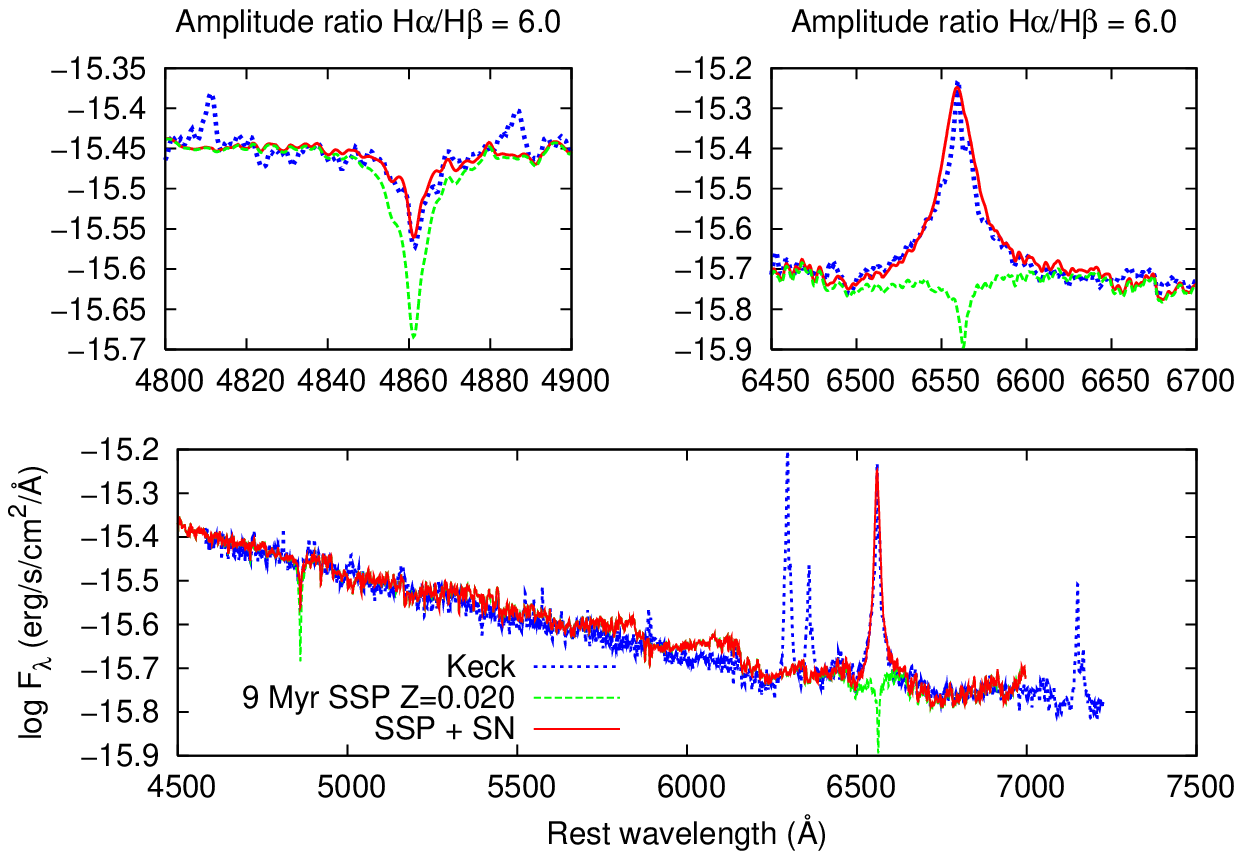}
\caption{The fitting of the deredshifted Keck spectrum 
with high-resolution SSP models having $Z=0.02$ (see text).
Each panel shows the observed Keck spectrum (blue), 
the best-fitting SSP spectrum (green), and the best-fitting 
SSP spectrum with the assumed H$\alpha$
and H$\beta$ SN emission lines added (red). The upper, 
smaller panels zoom in on the H$\beta$ and H$\alpha$ regions. The
assumed amplitude ratio is indicated at the top of the figures.}
\label{fig-hires}
\end{figure*}

The fitting was computed so that the sum of the SSP model spectrum 
and the adopted H$\beta$ emission profile with fixed AR
and $\gamma$ was fitted to the observed spectrum by varying the
age and the cluster mass. $E(B-V)$ was fixed at $0.1$ mag and
the model spectra were reddened using the same Galactic
reddening law \citep{fitz07} as for the SED fitting.  
We restricted the computation of $\chi^2$ in the
vicinity of the H$\beta$ line, between 4820~\AA\ and 4910~\AA\ rest
wavelengths. 

It was found that the H$\beta$ profile can be fitted
satisfactorily with a broad range of ages, depending on the chosen
metallicity and H$\alpha$/H$\beta$ AR. 
Figure~\ref{fig-hires} shows two of the best-fitting models
with $Z=0.02$ and AR$=4$ (left figure) and $6$ (right figure). 
The corresponding cluster ages are 25 Myr and 9 Myr, respectively. 
Assuming AR$=4$, the ages of the best-fitting models were found 
to be between 25 and 40 Myr depending on metallicity. 
However, they turned out to be 
30 -- 45 Myr for AR$=3$, and 9 -- 25 Myr for AR$=6$. 
The results from SSP models based on Padova evolutionary tracks 
\citep{del05} showed very similar behavior, but resulted in 
higher cluster ages and poorer fits (i.e. higher $\chi^2$).

It is concluded that in general, the fitting of the high-resolution 
spectrum confirms that the cluster is probably younger than 
50 Myr, but the H$\beta$ profile turned out to be mostly sensitive 
to the non-negligible contamination from SN~2004dj. 
As a result, the line-profile analysis could not lead
to a unique solution for the cluster age and metallicity. Hence, at
present, we cannot use the H$\beta$ line-profile analysis to further 
constrain the cluster age.

\begin{figure}
\plotone{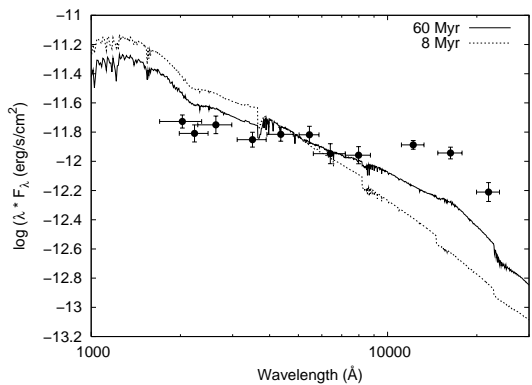}
\caption{Comparison of continuous SFR models with the observed SED of
S96.  The models were generated by the Starburst99 code based on
Padova AGB-enhanced tracks with $Z=0.008$ and Kroupa IMF.  The ages of
the models are indicated in the legend.}
\label{fig-sedfit3}
\end{figure}

We have also examined the hypothesis that S96 may not be a SSP
resulting from a single, rapid initial starburst.  Although a single
starburst is a more plausible mechanism for the formation of a
massive, compact stellar cluster, continuous star formation is taking
place within the disk of NGC~2403 \citep{dav07}.  We have checked
whether the SED of S96 could be fitted by that of a SSP resulting from
continuous star-formation rate (SFR); Starburst99 models were compared
with Geneva tracks, Kroupa IMF, and different metallicities assuming
continuous SFR.  The SFR was simply scaled to match the $V$-band
observed flux of the cluster SED.  Two of the models with $Z=0.008$
metallicity are plotted in Figure~\ref{fig-sedfit3}. Regardless of
age, these models are too bright in the UV and too faint in the NIR,
which suggests that the observed SED cannot be described by continuous
SFR. The same result has also been obtained using other metallicities,
or applying the Padova evolutionary tracks. Note that the presence of
a hypothetical dense, intracluster dust cloud may significantly alter
the shape of the resulting SED, but a detailed study of such a model
would require much better observational coverage of S96 at IR
wavelengths.

\subsection{Isochrone Fitting}

The computed photometry of the $HST$/ACS frames (\S 2.1.3) was used to
construct color-magnitude diagrams (CMDs) of S96 using either $B-V$ or
$V-I$ as color. We have selected and examined all resolved stars
within $R = 35$ pixels ($\sim 15$ pc) around the cluster center (green
circle in Fig.~\ref{fig-hst}) as possible cluster members.  Note that
the visible diameter of the unresolved inner part of the cluster is
$\sim 15$ pixels, corresponding to $\sim 6$ pc at the distance of
NGC~2403.

\begin{figure*}
\epsscale{1.0}
\plottwo{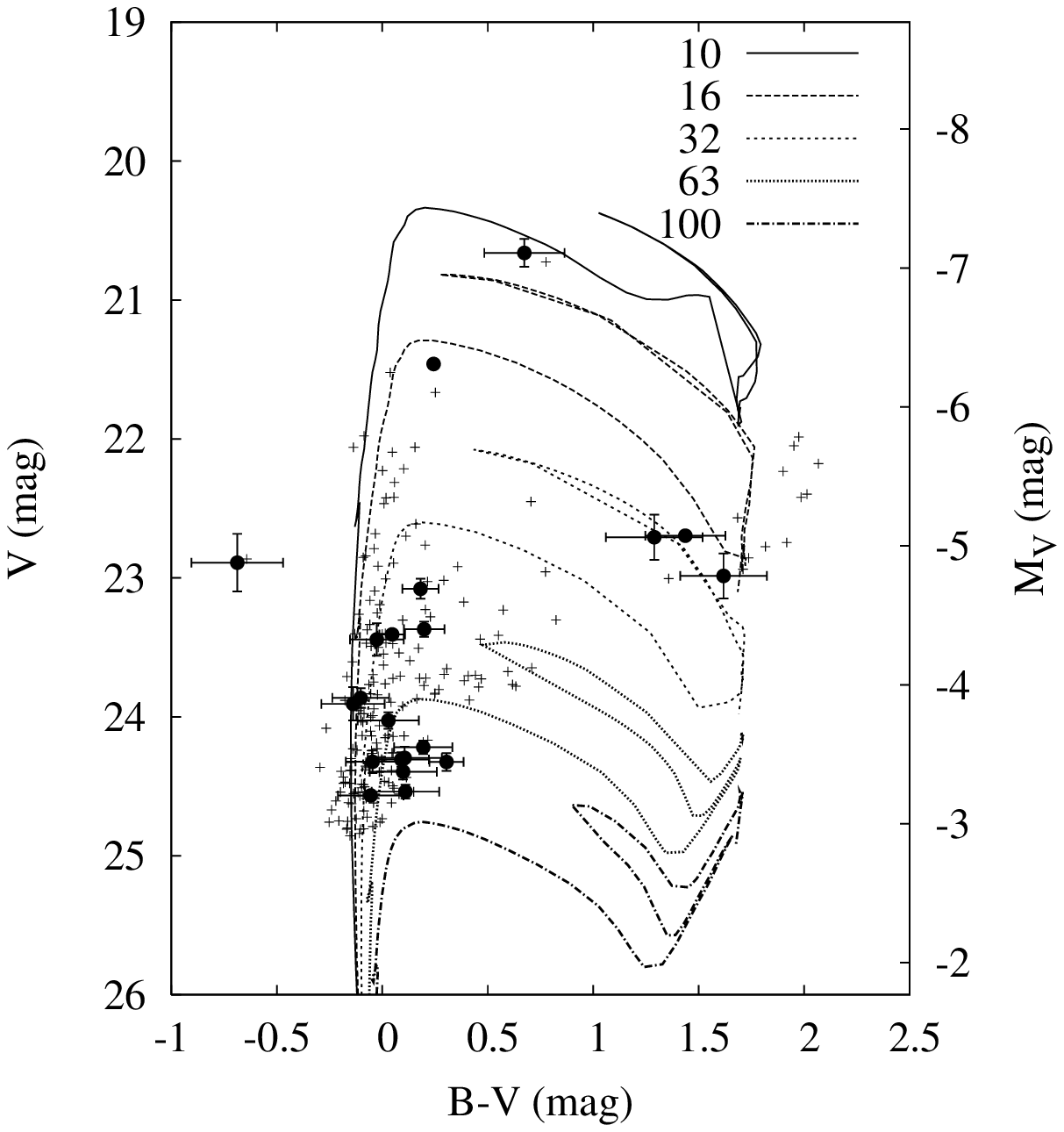}{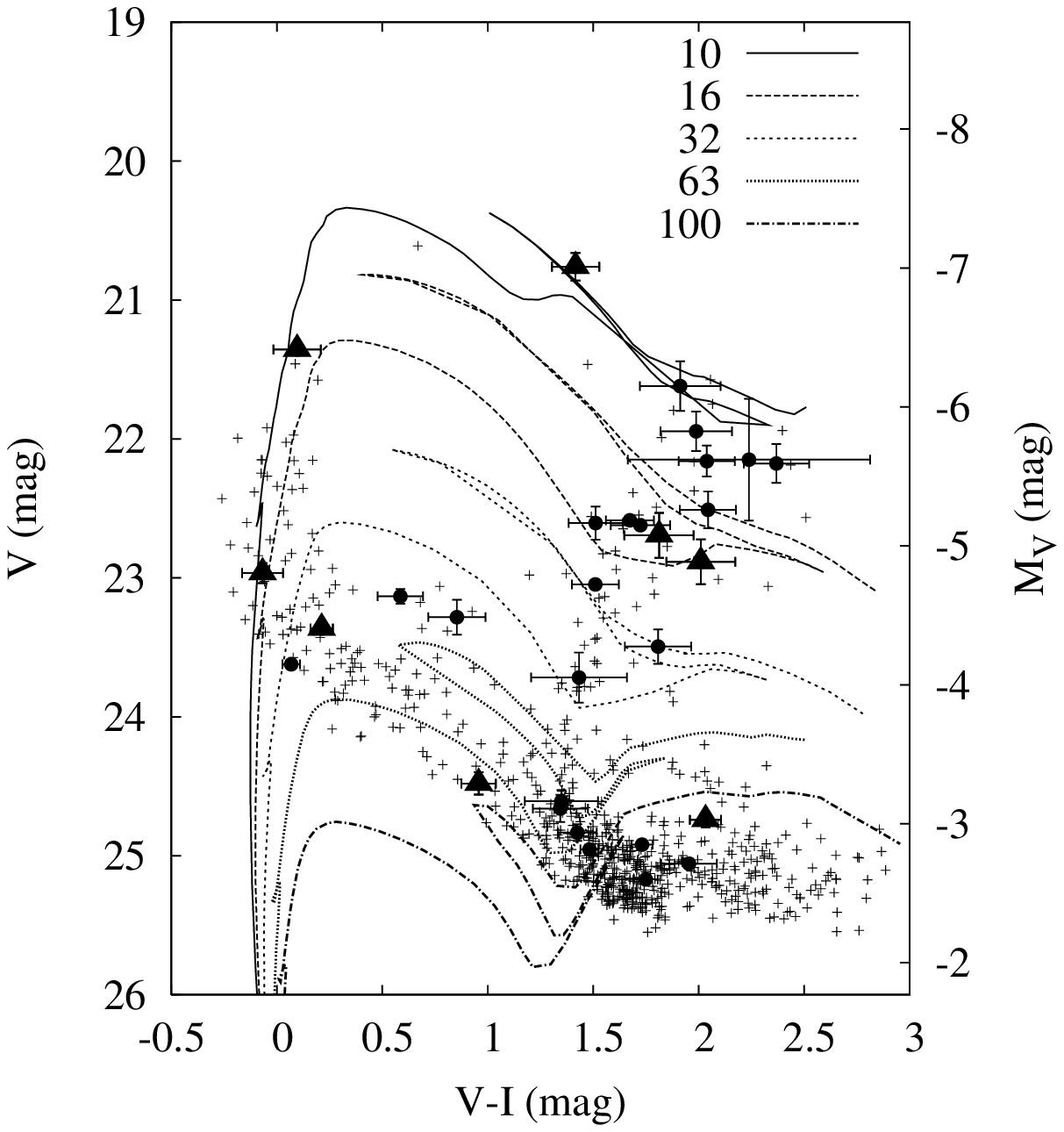}
\caption{$HST$ color-magnitude diagrams of S96 using $B-V$ (left panel)
and $V-I$ (right panel) colors. Filled symbols represent the cluster 
stars (within the encircled region in Fig.~5), while the plus signs 
denote field stars. 
In the right panel, those cluster stars that have
both $B-V$ and $V-I$ colors are plotted as triangles. 
The fitted Padova isochrones for $Z=0.019$ are 
also shown (labels indicate the cluster age in Myr). }
\label{fig-cmd}
\end{figure*}

The CMDs are plotted in Figure~\ref{fig-cmd}, where the filled circles
denote the possible cluster members, within the $R = 35$ pixels radius
(referred to as the ``cluster region'' hereafter), 
while crosses represents the other field stars. outside the cluster region. 

Thirty stars have a measured $V-I$ color within the 
cluster region. There are 21 such stars with $B-V$ color.
However, only 7 stars are common to the two samples, 
due to the reduced sensitivity of ACS in the blue. 

The field-star contamination within this region was estimated by
putting outside the cluster region an annulus having the same area as
that of the cluster region, and counting the stars within this
annulus. Using different inner radii for the annulus, but keeping its
area fixed, the number of field stars was found to vary between	 $1$
and $5$. Adopting its mean value, the expected number of field stars
within the cluster region is $3 \pm 2$. The relative contamination of
projected field stars within the cluster region is $\sim 10$\%.
Assuming that the positions of field stars follow a Poisson
distribution with $\lambda=3$ as the expected value, the probability
of the occurrence of 8 field stars within the cluster region
(i.e., $\sim 26$\% contamination) is $\sim 0.8$\%.  This
number strongly suggests a 99\% probability that at
least 22 stars found within the cluster region are indeed physically
associated with S96, and not just a random concentration of unrelated
field stars. 

\begin{figure}
\plotone{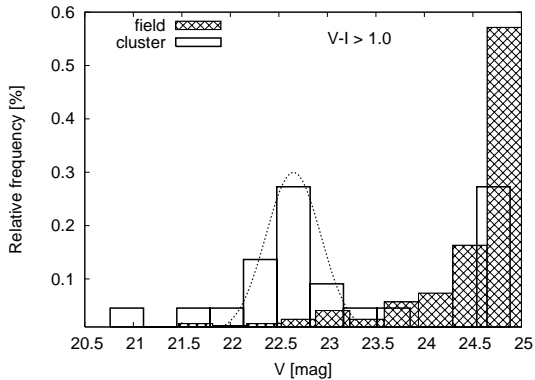}
\caption{Relative frequency of field stars (filled bars) and cluster stars (open bars)
with $V-I > 1$ mag as a function of observed $V$ magnitude. The dotted
curve is a Gaussian illustrating the presence of the bright cluster stars
at $V \approx 22.5 \pm 0.5$ mag. }
\label{fig-redhist}
\end{figure}

The separation of the cluster members and the field stars can also be
illustrated in their magnitude histogram.  In Figure~\ref{fig-redhist}
the relative frequency (i.e., the number of stars in a magnitude bin
divided by their total number) of the field stars (filled bars) and
those within the cluster area (open bars) that have $V-I > 1$ mag is
plotted as a function of the observed $V$ magnitude.  The distribution
of these red stars clearly indicates that in the cluster area there is
a significant excess of stars at $V \approx 22.5 \pm 0.5$ mag. Their
magnitude distribution can be roughly approximated by a Gaussian, and
it is markedly different from that of the field stars, being
monotonically increasing toward fainter magnitudes.

Turning back to Figure~\ref{fig-cmd}, it also contains the latest
Padova isochrones \citep{cio06a, cio06b} including variable molecular
opacities in the thermally pulsing asymptotic giant branch (TP-AGB)
phase, assuming solar metallicity. The ages of the plotted isochrones
(10, 16, 32, 63, and 100 Myr) are indicated in the legend.  The
isochrones were reddened with $E(B-V) = 0.1$ mag (\S 3.1) assuming the
Galactic reddening law \citep{fitz07}, and shifted to the $3.5$ Mpc
distance of the host galaxy.  This reddening value seems to be a good
estimate for the other field stars as well.  The $Z=0.019$ tracks were
selected, because the fitting of the integrated cluster SED produced
the best results using this metallicity (see \S 3.1). Comparing the
CMDs with isochrones of $Z=0.008$ and $0.004$, it was found that these
isochrones do not extend enough to the red (to $V-I \approx 2$ mag)
where some of the bright cluster stars reside. However, the age
distribution of the observed stars (i.e., the concentration of stars
along the computed isochrones) is the same as in the case of
$Z=0.019$, so the age limits of the resolved population of S96 are
found to be rather insensitive to the actual metallicity of the
cluster.

It is interesting that in the CMDs the field stars follow roughly the
same distribution as the cluster members.  Note that the blueward
distribution of all stars in the $V$ vs. $B-V$ diagram below $23$ mag,
and the redward distribution in the other CMD below $23.5$ mag, are
due to the decreasing sensitivity of the detector/filter combination
in that color regime (that is, the incomplete detection of
objects). The completeness limit (the magnitude limit above which all
stars are detected regardless of their color) was estimated as $V
\approx 22.5$ ($M_V \approx -5.2$) mag for the $V$ vs. $B-V$ diagram,
and $V \approx 23.5$ ($M_V \approx -4.2$) mag for the $V$ vs. $V-I$
diagram.  In order to have better statistics, in the following we
analyze the $V$ vs. $V-I$ diagram.

From Figure~\ref{fig-cmd} the age of each cluster star was determined
as the age of the nearest isochrone.  In some cases, when different
isochrones ran very close to each other, only upper and lower limits
(e.g., $63 < t < 100$ Myr) could be determined.

It is apparent that the brightest cluster stars are closest to the
$\sim 10$ Myr isochrone consistently in both diagrams.  However, there
are only 2 or 3 such stars, so they may also be binaries consisting of
older/fainter stars.  Most of the bright resolved stars have $V-I
\approx 2$ mag and are distributed between the $10$ and $16$ Myr
isochrones. These are in very good agreement with the ages of the
SED fitting with the lowest $\chi^2$ (\S 3.1). Because these stars are
expected to have the most significant contribution to the integrated
cluster SED, this agreement gives further credibility to the age
estimates found in \S 3.1.

On the other hand, 16 cluster stars out of 30 ($\sim$50\% of the
resolved cluster population) are close to or below the 32~Myr
isochrone.  The detection becomes increasingly color dependent below
23 mag, so the actual number of such stars may be higher. There are a
few very red stars at $\sim$25 mag, where we cut the observed sample,
because the errors calculated by DOLPHOT started to exceed 1 mag (note
that the real brightness uncertainties of these stars may be higher,
but we used the errors given by DOLPHOT as a selection criterion).

\begin{figure}
\plotone{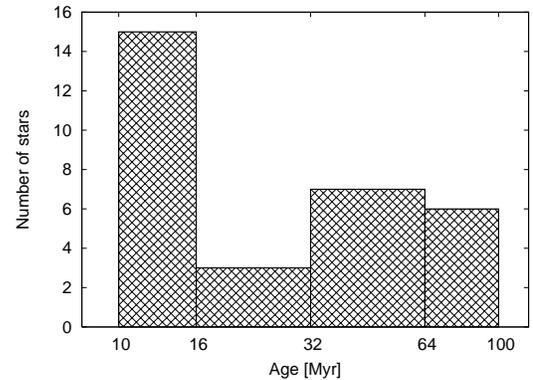}
\caption{The histogram of the computed ages of cluster stars.}
\label{fig-agehist}
\end{figure}

Fig.\ref{fig-agehist} shows the histogram of the ages (the age
resolution follows that of the isochrones).  About half of the
resolved cluster members fall into the 10--16 Myr age interval,
while the other 50\% have ages distributed between 16 and 100
Myr.  These results suggest that the resolved population of S96 cannot
be represented by a single age.  Instead, a ``young'' population with
an age of 10--16 Myr and an ``old'' population at 30--100 Myr
seem to exist within the cluster area. 

\begin{figure}
\plotone{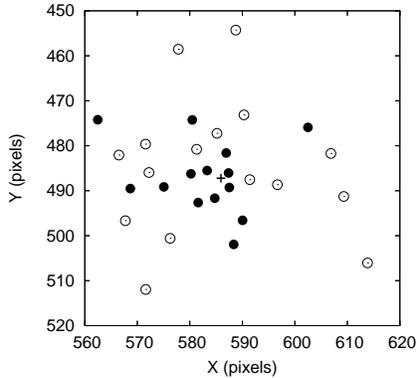}
\caption{The spatial distribution of ``young'' stars (filled
circles) and ``old'' stars (open symbols) within the cluster
region. The approximate center of the cluster is marked with a
``+'' sign. }
\label{fig-xy}
\end{figure}

\begin{figure*}
\plottwo{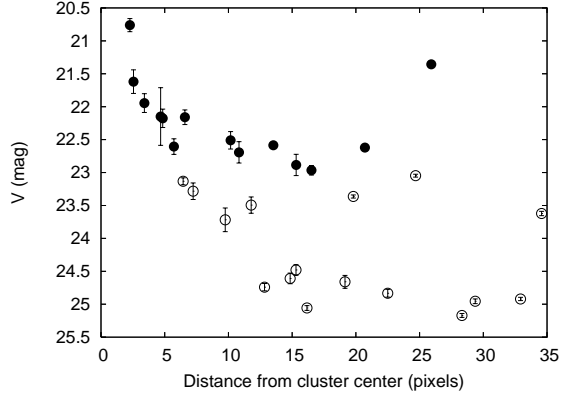}{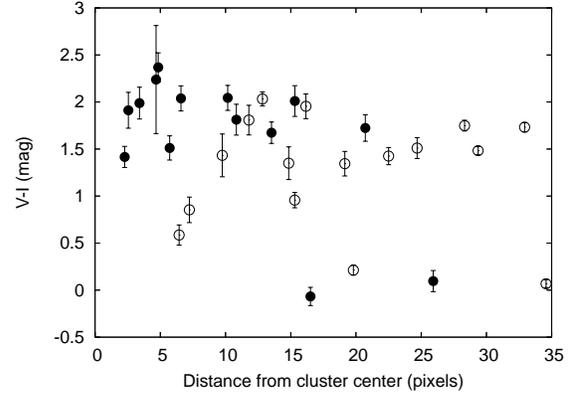}
\caption{$V$ magnitudes (left panel) and colors
(right panel) of the ``young''
and ``old'' members of S96 as a function of
their distance from the cluster center. The symbols
have the same meaning as in Fig.~\ref{fig-xy}.}
\label{fig-rad}
\end{figure*}

It is interesting to compare the spatial distribution of the ``young''
and ``old'' stars in the cluster area. This is shown in
Figure~\ref{fig-xy}, where the image coordinates of the resolved stars
(in pixels) are plotted. Filled circles represent the ``young'' stars,
and open symbols denote the ``old'' ones. The ``young'' (bright) stars
appear to be concentrated around the cluster center, while the ``old''
stars are more scattered. This is illustrated further in
Figure~\ref{fig-rad}, where the $V$ magnitudes and the $V-I$ colors
are plotted as a function of the radial distance from the cluster
center (in pixels).  Again, the ``young'' stars seem to dominate the
inner area within $r \approx 6$ pixels.  On the other hand, the light
from the unresolved part of the cluster is also strong here, making
the detection of fainter stars very difficult in the central area.
Thus, the lack of ``old'' stars in this area is surely affected by
selection. The right-hand panel of Figure~\ref{fig-rad} suggests that
the average color of the ``young'' population is somewhat redder than
that of the ``old'' population.

Summarizing the results obtained in this section, we conclude that the
resolved stars in S96 belong to two populations having ages of 10--16
Myr and 30--100 Myr.  The ``young'' stars are brighter and somewhat
redder that the ``old'' ones, and they are located closer to the
central part of the cluster. The age limits of the ``young''
population are in good agreement with the lowest $\chi^2$ models in the
SED fitting (\S 3.1), when the random population of the IMF was taken
into account, but slightly higher than the age of the best-fitting
canonical models ($\sim 8$ Myr), i.e.,  those without random IMF
population. 

\subsection{The Absence of H$\alpha$ Emission Around S96}

\begin{figure}
\plotone{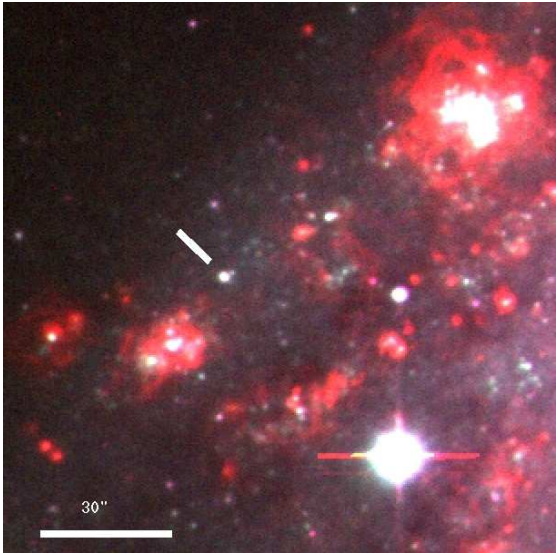}
\caption{False-color image of NGC 2403 in the vicinity of S96 
(red, H$\alpha$; green, $V$; blue, $B$) obtained with the 
2.3~m Bok telescope at Steward Observatory. The field of
view is about $2' \times 2'$; north is up and east to the left. 
The marked object is S96. The extended red areas are nearby
\ion{H}{2} regions.}
\label{fig-ha}
\end{figure}

NGC~2403 is known to show intense star-forming activity \citep{dav07}.
From deep $gri$ and $JHK$ imaging, \citet{dav07} found that the SFR
during the past 10 Myr has been $\sim 1$ M$_\odot$ yr$^{-1}$ in the
whole disk of NGC~2403.  The SFR was highest in the region at
galactocentric distances 2--4~kpc.  The intense star formation in the
inner disk may explain the existence of young ($\sim 8$--10 Myr)
compact clusters, such as S96 which is at $R_{GC} \approx 2.7$ kpc.

Young clusters are able to ionize the surrounding hydrogen clouds,
showing up as large, bright \ion{H}{2} regions. The measured H$\alpha$
luminosity is known to correlate with the SFR of these complexes
\citep{kenn98, pfl07}. The ionizing UV photons come mostly from the
young, massive OB stars located inside the clouds. Because the
lifetime of such stars is short, the number of ionizing photons
decreases rapidly after $\sim$7--8 Myr for clusters/associations that
were formed after an initial starburst \citep{dopita06}. Thus, the
presence/absence of H$\alpha$ emission around S96 may give an
additional, independent constraint on the age of the cluster.

Figure~\ref{fig-ha} shows the color-combined image of NGC~2403
obtained with the 2.3~m Bok telescope at Steward Observatory (see \S
2) using $B$, $V$, and H$\alpha$ filters for the blue, green, and red
colors, respectively. It is apparent that there are a number of
extended \ion{H}{2} regions showing H$\alpha$ emission in the vicinity
of S96 (the marked object), as expected in a stellar field with
ongoing star formation. Following the method applied recently by
\citet{ramy07}, the SFRs of these complexes were estimated to be
0.01--0.001 M$_\odot$ yr$^{-1}$, typical of such H$\alpha$-emitting
regions.  However, S96 appears stellar, without any indication for
extended H$\alpha$ emission. This suggests that the flux at H$\alpha$
is coming entirely from inside the unresolved cluster.  Indeed, it is
very likely that the source of this emission is mostly from SN~2004dj
(\S 2.1.2).

The lack of any extended H$\alpha$ emission around S96 can be used to
estimate a lower limit for the cluster age, as outlined above. The
number of ionizing UV photons as a function of age was estimated by
the Starburst99 code (see \S 3.1) applying Geneva tracks, Salpeter
IMF (but neglecting random IMF sampling), and $Z=0.02$. The
cluster mass was fixed at $M_c = 50,000$ M$_\odot$, between
the cluster masses derived during the SED fitting (see
Table~\ref{tbl-sedfit1}). The calculated numbers of ionizing photons
have been converted to the radius of the \ion{H}{2} region
applying the formula 
\begin{equation} R_{\rm H~II} = {3 \over {4 \pi}} 
{Q(H^0) \over {N_e^2 \alpha_B}}, 
\end{equation} 
\noindent
where $Q(H^0)$ is the number of photons capable of ionizing hydrogen,
$N_e$ is the number density of electrons (complete ionization was
assumed: $N_e = N_p \approx N_H$), and $\alpha_B$ is the effective
recombination coefficient for H \citep{osterbrock}. The value of
$\alpha_B$ was estimated using 
\begin{equation} \alpha_B = 2.59 10^{-13} \left({T_e \over 10^4} 
\right)^{-0.833} 
\end{equation}
assuming $T_e = 10^4$ K \citep{moore02}.

\begin{figure}
\plotone{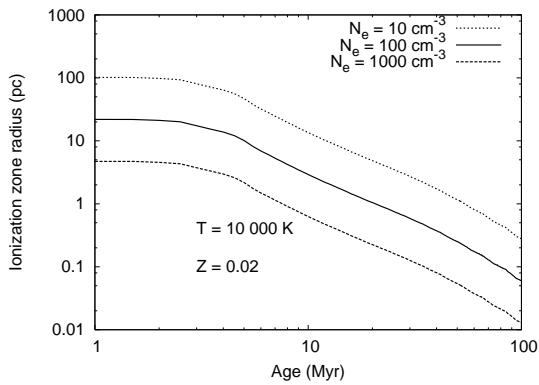}
\caption{The calculated radius of the ionization zone (in pc) 
as a function of cluster age (in Myr). The continuous line corresponds
to $N_e = 100$ cm$^{-3}$, while the dashed and
dotted lines illustrate the dependence of the result on this parameter. }
\label{fig-ionz}
\end{figure}

\begin{figure}
\plotone{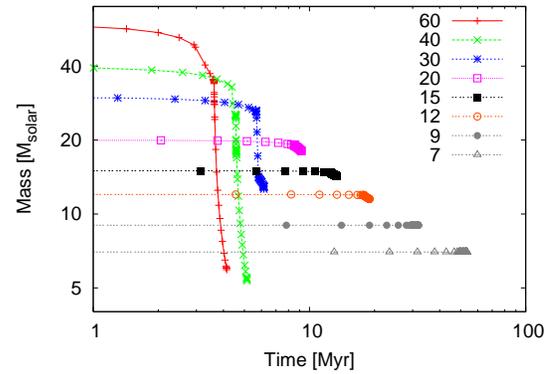}
\caption{Time dependence of masses of stars with $M > 7$~M$_\odot$ 
from Padova tracks. Labels indicate the mass in M$_\odot$. Each curve
ends at the age of the last model of the corresponding track, indicating
the lifetime of a star with the given initial mass.}
\label{fig-mass}
\end{figure}

\begin{figure}[h]
\plotone{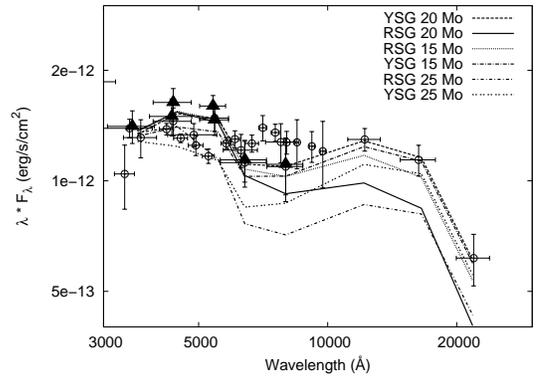}
\caption{Comparison of pre-explosion (open circles) and
post-explosion (filled triangles) cluster SED data.  Different lines
show the change of the SED if the SN explosion destroyed a yellow
supergiant (YSG) or a red supergiant (RSG) with a mass indicated
by the labels.}
\label{fig-sedcomp}
\end{figure}

In Figure~\ref{fig-ionz} the radius of the ionization zone is plotted as 
a function of the cluster age. The continuous line shows the results
for $N_e = 100$ cm$^{-3}$ (a typical electron density in bright
\ion{H}{2} regions), while the dashed and dotted lines illustrate
the results if $N_e$ was an order of magnitude higher or lower.
Note that changing the cluster metallicity down to $Z = 0.004$ 
caused only negligible alterations in these curves. 
It is apparent that at $\sim 10$ Myr the ionized cloud
has $\sim 5$ pc radius, which is similar to the radius of S96 as
seen by $HST/ACS$ (\S 2.1.3). Above 10 Myr the radius
quickly decreases. At $\sim 20$ Myr it is only $\sim 1$ pc,
which is much less than the size of the cluster. This suggests
that the $10 < t_c < 20$ Myr cluster age found in the previous
sections is consistent with the lack of resolved \ion{H}{2}
region around S96. 

For an age of $\sim 8$ Myr, which was proposed by the fitting of SEDs
without random IMF fluctuations, $R_{\rm H~II} \approx 10$ pc, which
is slightly larger than the visible cluster size.  Thus, the $\sim 8$
Myr age may be less probable than the $t_c \geq 10$ Myr ages found
above. However, if $N_e > 100$ cm$^{-3}$ is allowed, $R_{\rm H~II}$
can be easily reduced to $\sim 5$ pc at $\sim 8$ Myr.

It is concluded that using the absence of an H$\alpha$-emitting region
around S96 results in a lower limit of the cluster age of $\sim$8--10
Myr. This is consistent with the age estimates of S96 found in the
previous sections.

\section{Discussion}

In \S 3.1, \S 3.2, and \S 3.3, constraints on the age of S96 were
derived with different techniques. The fitting of theoretical SEDs (\S
3.1) gave possible ages distributed between 8 Myr and 40 Myr depending
on the cluster metallicity and the models applied. The most probable
solutions turned out to be between 10 and 25 Myr.

The fitting of isochrones to the CMDs of the resolved stellar
population in the vicinity of S96 (\S 3.2) resulted in two distinct
populations with ages of 10--16 Myr and 30--100 Myr. The younger stars
seem to be somewhat redder, and they are located closer to the cluster
center than the members of the older population.

The absence of an \ion{H}{2} region emitting in H$\alpha$ around S96
is consistent with the lower age limit of $\sim 10$ Myr. As the
simulations with Starburst99 indicate (\S 3.3), the predicted radii of
such a cloud after $t > 10$ Myr would decrease below $\sim 5$ pc,
roughly the projected radius of S96.  

How can we explain the existence of populations with two different
ages within such a compact cluster?  The most likely hypothesis is the
capture of field stars by the massive stellar cluster during its
formation, as discussed recently by \citet{pflk07a} for explaining the
existence of stars with $t \approx 10$--18 Myr within the Orion
Nebula cluster, where most stars have $t<3$ Myr. The age discrepancy
is similar to the case of S96, but otherwise the situation is
different, because S96 is much more massive than the Orion Nebula
cluster, and the older population resides in the outer region of S96.

Following the argument of \citet{pflk07a}, the collapsing pre-cluster
cloud may capture nearby field stars during its collapse time, which
is roughly equal to its free-fall timescale, $\tau_{ff} \approx (R_c^3
/ G M_c)^{1/2}$, where $R_c$ is the initial radius of the cloud at the
start of the collapse and $M_c$ is the total mass of the
cloud. Adopting $R_c \approx 15$--20 pc and $M_c \approx$ ~(25--100)
$\times 10^3$~M$_\odot$, the estimated collapse time of S96 is $\tau_c
\approx 2$--8 Myr. Note that this is an upper limit for the collapse
time, since the mass of the cloud can be significantly higher than the
final stellar mass of the cluster (which was used to estimate $M_c$).
Assuming the number density of stars as $2.44$ pc$^{-3}$,
\citet{pflk07a} calculated the number of captured field stars within
2.5 pc of the Orion Nebula cluster center as $100 < N < 1000$ if
$\tau_c > 2$ Myr. These results suggest that for S96, which is an
order of magnitude more massive than the Orion Nebula cluster, the
number of captured field stars should be substantial, even if the
number density of the surrounding stars is lower than that around
the Orion Nebula cluster.  Moreover, we have studied the stellar
content within $\sim 15$ pc from the center of S96, instead of 2.5 pc,
which may also increase the number of captured stars.

If we interpret the higher age of the stars in the outer part of S96
as a result of field-star capture, the photometric data suggest that
most of the stars resolved by ACS are captured, and the young ($t
\approx 10$--16 Myr), brightest, most massive stars that mostly
determine the shape of the integrated SED reside within the inner ($R
\approx 3$ pc), unresolved cluster core.  This configuration is
roughly consistent with that of other young clusters, as the
brightest, most massive members are generally found closest to the
center.

There might be other mechanisms responsible for the age dispersion
within clusters.  A possible hypothesis could be continuous star
formation within a $T_c \approx 60$ Myr cloud.  This scenario can
certainly be ruled out, because the SED of such a stellar population
is not compatible with the observations (see Fig.~\ref{fig-sedfit3}).
However, as also suggested by the referee, there might be a rapid
initial starburst that occurred within a region having continuous star
formation.  In this case the resulting SED would be a flux-weighted
combination of the SEDs from a continuous SFR (Fig.~\ref{fig-sedfit3})
and from a starburst (Fig.~\ref{fig-sedfit}).  While this scenario
would certainly be worth study in detail, the construction of such
customized models is beyond the scope of this paper.

\citet{pflk07b} discuss yet another possibility: gas accretion (and
subsequent star formation) from the nearby interstellar medium (ISM)
by a massive cluster. However, according to their simulations, this
process is expected to work only for $M_c > 10^6$~M$_\odot$ cluster
masses and have a characteristic timescale of a few Gyr. Thus, it is
probably insignificant for S96, for which both the cluster mass and
the considered timescale are an order of magnitude less.

The age of S96 is a key parameter in constraining the mass of the
progenitor of SN~2004dj. The classical theoretical lower limit for the
collapse of a stellar core is $\sim 8 ~M_\odot$, but this can be 1 --
2~$M_\odot$ smaller depending on the treatment of core convective
overshooting \citep{woo02}. Recent direct identifications of Type II-P
supernova progenitors typically have masses of $\sim 8$--15 M$_\odot$
\citep{ma05, li06}, and none of them clearly exceed $M \approx
20$~M$_\odot$.

The fact that SN~2004dj occurred close to the projected center of S96
(\S 2.1.3 and Fig.~\ref{fig-hst}) strongly suggests that its
progenitor was indeed a cluster member. Although S96 may contain a
significant number of older stars captured from the field, it is more
probable that a $M \geq 7~M_\odot$ star is formed during or after the
collapse of the pre-cluster cloud.  Assuming this scenario, the $M
\geq 7~M_\odot$ limit implies $t \leq 60$ Myr as an upper limit for
the cluster, according to Padova isochrones. This is in good agreement
with the ages of most of the resolved cluster stars inferred from
isochrones, because even the members of the older population have ages
comparable to or less than 60 Myr.

In Figure~\ref{fig-mass} the masses of $M \geq 7~M_\odot$ stars are
plotted as a function of age from the same Padova evolutionary tracks
as above.  The final ages of the curves correspond to the last
theoretical model for a given initial mass.  Note that the Padova
evolutionary tracks do not extend up to the actual moment of core
collapse, so the final ages for all masses are only lower limits, but
an age excess as large as $\sim 10$\% is hardly expected. If we accept
the $\sim 10$ Myr age for S96 as a lower limit inferred from both SED
fitting and isochrones, this would imply $M_{\rm prog} \approx
20$~M$_\odot$ for the initial mass of the progenitor.

From fitting the pre-explosion SED, \citet{maiz04} and \citet{wang05}
estimated $M_{\rm prog} \approx 12$--15~M$_\odot$, which agrees very
well with the most probable age of 10--20 Myr found in \S 3.1 and 3.2.
On the other hand, \citet{v06} obtained a significantly lower age and
higher progenitor mass ($\geq 20~M_\odot$) from nearly the same
observed data as \citet{maiz04}, but using different model SEDs.  This
would require $T_c \approx 8$ Myr, which is lower than most of the age
estimates discussed above, but may not be ruled out entirely, because
certain SED models indeed predict such young age.  However, these
earlier results were more affected by the age-reddening degeneracy
(see \S 3.1), because of the restricted wavelength range of the
observed SED.  

There is yet another way to test the possible mass of the progenitor
star via the effect of the SN explosion on the integrated cluster SED,
as first suggested by \citet{maiz04}.  The explosion of SN~2004dj must
have changed slightly the supergiant population of S96, because one
bright (perhaps the brightest) star was missing after the SN faded
away. This should be apparent in the cluster SED as well, altering
both the overall flux level as well as the spectral shape of the
post-explosion SED.  The difference between the pre- and
post-explosion SED is approximately the flux spectrum of the
progenitor star just before explosion. If the progenitor is a red
supergiant (RSG), then mostly the NIR region of the cluster SED will
be depressed, while if it is a yellow supergiant (YSG), the change
will be more pronounced in the optical.

Figure~\ref{fig-sedcomp} shows a comparison of the observed pre- and
post-explosion cluster SEDs with the predictions of this
hypothesis. The lines represent the theoretical post-explosion cluster
SED if a RSG or a YGS with a given mass is removed from the
pre-explosion SED. $M_{\rm prog} = 25$, 20, and 15~M$_\odot$ were
selected, and their fluxes at different bands were determined from the
Padova evolutionary tracks assuming $Z=0.02$.

It is apparent that the removal of a $M_{\rm prog} = 25 ~M_\odot$ star
would cause a strong flux decrease above 5000~\AA\ that clearly
exceeds the uncertainty of the observed flux levels. Such a massive
progenitor is therefore unlikely. The lack of a 20~M$_\odot$ RSG would
also result in a similar flux depression in the NIR. Unfortunately, no
post-explosion NIR photometry is at our disposal, so we could not
verify this prediction.  All of the other proposed progenitors do not
cause the flux to drop significantly below the observed SED.  

However, as our Keck spectrum suggests (\S 2.1.2), 
the nebular emission from the SN~2004dj ejecta may also have a
non-negligible contribution to the observed fluxes in the $R$ 
and $I$ bands via emission lines from H$\alpha$, 
[\ion{O}{1}] $\lambda \lambda$6300,6364, 
[\ion{Fe}{2}] $\lambda$7155 and possibly 
[\ion{Ca}{2}] $\lambda \lambda$7291,7324 
\citep{sahu06}. 
Also, the lack of post-explosion
observations in the NIR SED makes the comparison between observations
and model predictions uncertain at present.  
More observations, especially in the
$JHK$ bands, would be very useful to clarify this issue.  From
Figure~\ref{fig-sedcomp}, it seems that the progenitor mass was
probably $<25$~M$_\odot$, but a 15~M$_\odot$ or even a 20~M$_\odot$
star is a possible candidate.

Putting together all available information, we conclude that the new
multi-wavelength observations favor a progenitor star with $12
\lesssim M_{\rm prog} \lesssim 20$ M$_\odot$, if the progenitor was a
member of the younger population within S96. This is consistent with,
and perhaps somewhat higher than, typical measured SN~II-P progenitor
masses.  However, because of the presence of stars belonging to an
older ($\sim 60$ Myr) population within S96, it cannot be ruled out
that the progenitor was one of them, which would imply $M_{\rm prog}
\approx 7$--8 M$_\odot$, close to the lower limit for such SNe.

\section{Conclusions}

We have presented late-time photometry of SN~2004dj and the
surrounding cluster S96, extending the time coverage of the
observational sample up to $\sim 1000$ d after explosion. In the
optical, the continuum flux from SN~2004dj faded below the level of
the integrated flux of S96 in 2006 Sep., $\sim 800$ d after explosion.
The pre- and post-explosion SEDs of S96 show no significant
differences in the range 2000--9000~\AA. The nebular spectrum of
SN~2004dj at $\sim900$ d after explosion was dominated by the blue
continuum from S96 shortward of 6000~\AA, and by strong H$\alpha$,
[\ion{O}{1}] $\lambda\lambda$6300, 6363, and [\ion{Fe}{2}]
$\lambda7155$ emission line, characteristic of a typical nebular
spectrum of a SN~II-P.

We have examined the multi-wavelength observations of S96 by different
methods, in order to derive constraints on the cluster age and
evolutionary status.  The fitting of the cluster SED (using the
average of pre- and post-explosion fluxes) results in cluster ages
distributed between $\sim 8$ and $\sim 40$ Myr, with the best-fitting
solutions being within 10--20 Myr.  The observed reddening is $E(B-V)
\approx 0.10 \pm 0.05$ mag; its uncertainty is greatly reduced
compared with previous studies, due to the inclusion of the UV fluxes
from {\it Swift} and {\it XMM-Newton}.
 
S96 appears to be partly resolved in images obtained with $HST$/ACS on
2005 August 28 ($\sim 425$~d after explosion), although the light from
SN~2004dj was still very strong at that time.  We have computed
photometry of the ACS images obtained through the $F435W$, $F606W$,
and $F814W$ filters, and combined the magnitudes of the detected
stellar sources in color-magnitude diagrams. Theoretical isochrones
fitted to the observed CMDs reveal that the resolved stars in the
outskirts of the cluster have a bimodal age distribution. The younger
population consists of stars with ages of $10 < t < 16$ Myr, while the
members of the older one have $30 < t < 100$ Myr. The ages of the
older population has a distribution that is similar to that of the
field stars, not associated with S96. This similarity may suggest that
about half of the cluster stars resolved by ACS were captured from the
field population during the formation of S96.

The absence of a visible H$\alpha$-emitting cloud around S96 implies a
lower limit for the cluster age of $\sim$8--10 Myr, in agreement with
the other age estimates.

The 10 Myr age of S96 would imply a SN 2004dj progenitor mass of
$M_{\rm prog} \approx 20$~M$_\odot$, while the mass limit for core
collapse (7--8 M$_\odot$) would mean $t \approx 60$ Myr for the age of
the progenitor. This latter limit is consistent with the age of the
older population within S96, leaving the possibility of a low-mass
progenitor open. The age of the younger population (10--16 Myr)
corresponds to $M_{\rm prog} \approx$ 12--15 M$_\odot$, which seems to
be the most probable mass estimate at present. We verified that
even a 20~M$_\odot$ progenitor would be consistent with the
unobservable flux difference between the pre- and post-explosion
SEDs. However, more observations, especially in the $JHK$ bands,
would be essential to narrow the mass range of the progenitor.

\acknowledgments

This work was based in part on observations made with the NASA/ESA
{\it Hubble Space Telescope}, obtained from the Data Archive at the
Space Telescope Science Institute, which is operated by the
Association of Universities for Research in Astronomy, Inc., under
NASA contract NAS 5-26555.  It was partially supported by NASA grants
GO--10607 (B.S.) and GO--10182 (A.V.F.), by NSF grant AST--0607384
(A.V.F.), and by Hungarian OTKA grant TS049872 (J.V.).  We are
grateful for the support received from the {\sl Swift} Science Center.
An anonymous referee provided many useful suggestions and advice that
helped us extend and improve the paper.  The SIMBAD database at CDS,
the NASA ADS and NED, and the Canadian Astronomy Data Centre have been
used to access data and references.

{\it Facilities:} \facility{Bok}, \facility{{\it HST} (ACS)}, 
\facility{Keck:II (DEIMOS)}, \facility{{\it Swift} (UVOT)}, 
{\it XMM-Newton}, Konkoly


\begin{thebibliography}{}

\bibitem[Brown et al.(2007)]{brown07} Brown, P.~J., et al.\ 2007,
\apj, 659, 1488

\bibitem[Bruzual \& Charlot(2003)]{bruz03} Bruzual, G., \& Charlot,
S.\ 2003, \mnras, 344, 1000

\bibitem[Buzzoni et al.(2007)]{buz07} Buzzoni, A., Bertone, E.,
Chavez, M., \& Rodriguez-Merino, L.~H.\ 2007, ArXiv e-prints, 709,
arXiv:0709.2711

\bibitem[Cervi{\~n}o et al.(2002)]{cervi02} Cervi{\~n}o, M.,
Valls-Gabaud, D., Luridiana, V., \& Mas-Hesse, J.~M.\ 2002, \aap, 381,
51

\bibitem[Cervi{\~n}o \& Luridiana(2004)]{cervi04} Cervi{\~n}o, M., \&
Luridiana, V.\ 2004, \aap, 413, 145

\bibitem[Cervi{\~n}o \& Luridiana(2006)]{cervi06} Cervi{\~n}o, M., \&
Luridiana, V.\ 2006, \aap, 451, 475

\bibitem[Cioni et al.(2006a)]{cio06a} Cioni, M.-R.~L., Girardi, L.,
Marigo, P., \& Habing, H.~J.\ 2006a, \aap, 448, 77

\bibitem[Cioni et al.(2006b)]{cio06b} Cioni, M.-R.~L., Girardi, L.,
Marigo, P., \& Habing, H.~J.\ 2006b, \aap, 452, 195

\bibitem[Crockett et al.(2007a)]{cr07a} Crockett, R.~M., et al.\
2007a, \mnras, 381, 835

\bibitem[Crockett et al.(2008)]{cr07b} Crockett, R.~M., et al.\ 2008,
\apj, 672, 99

\bibitem[Davidge(2007)]{dav07} Davidge, T.~J.\ 2007, \apj, 664, 820

\bibitem[Dolphin(2000)]{dolph00} Dolphin, A.~E.\ 2000, \pasp, 112,
1383

\bibitem[Dopita et al.(2006)]{dopita06} Dopita, M.~A., et al. 2006,
\apj, 647, 244

\bibitem[Faber et al.(2003)]{Faber03} Faber, S.~M., et al. 2003,
\procspie, 4841, 1657

\bibitem[Fierro et al.(1986)]{fierro86} Fierro, J., Torres-Peimbert,
S., \& Peimbert, M.\ 1986, \pasp, 98, 1032

\bibitem[Filippenko(1982)]{fil82} Filippenko, A.~V. 1982, PASP, 94, 715

\bibitem[Filippenko(1997)]{fil97} Filippenko, A.~V. 1997, ARAA, 35,
309

\bibitem[Fitzpatrick \& Massa(2007)]{fitz07} Fitzpatrick, E.~L., \&
Massa, D.\ 2007, \apj, 663, 320

\bibitem[Gehrels et al.(2004)]{geh04} Gehrels, N., et al. 2004, \apj,
611, 1005

\bibitem[Gonz\'alez Delgado et al.(2005)]{del05} Gonz\'alez Delgado,
R.~M., Cervi{\~n}o, M., Martins, L.~P., Leitherer, C., \& Hauschildt,
P.~H.\ 2005, \mnras, 357, 945

\bibitem[Hendry et al.(2006)]{hen06} Hendry, M.~A., et al.\ 2006,
\mnras, 369, 1303

\bibitem[Immler et al.(2007)]{imm07} Immler, S., et al.\ 2007, \apj,
664, 435

\bibitem[Jamet et al.(2004)]{jamet04} Jamet, L., P{\'e}rez, E.,
Cervi{\~n}o, M., Stasi{\'n}ska, G., Gonz{\'a}lez Delgado, R.~M., \&
V{\'{\i}}lchez, J.~M.\ 2004, \aap, 426, 399

\bibitem[Jimenez et al.(2004)]{jim04} Jimenez, R., MacDonald, J.,
Dunlop, J.~S., Padoan, P., \& Peacock, J.~A.\ 2004, \mnras, 349, 240

\bibitem[Kaviraj et al.(2007)]{kav07} Kaviraj, S., Rey, 
S.-C., Rich, R.~M., Yoon, S.-J., \& Yi, S.~K.\ 2007, \mnras, 381, L74 

\bibitem[Kennicutt(1998)]{kenn98} Kennicutt, R.~C., Jr.\ 1998, 
\araa, 36, 189 

\bibitem[Koleva et al.(2008)]{koleva08} Koleva, M., Prugniel, P.,
Ocvirk, P., Le Borgne, D., \& Soubiran, C.\ 2008, \mnras 385, 1998

\bibitem[Larsen(1999)]{larsen99} Larsen, S.~S.\ 1999, \aaps, 139, 393

\bibitem[Leonard et al.(2008)]{leonard08} Leonard, D.~C., Gal-Yam, A.,
Fox, D. B., Cameron, P. B., Johansson, E. M., Kraus, A. L., Le Mignant,
D., \& van Dam, M. A.\ 2008, submitted (arXiv:0809.1881)

\bibitem[Li et al.(2006)]{li06} Li, W., Van Dyk, S.~D., Filippenko,
A.~V., Cuillandre, J.-C., Jha, S., Bloom, J.~S., Riess, A.~G., \&
Livio, M.\ 2006, \apj, 641, 1060

\bibitem[Li et al.(2007)]{li07} Li, W., Wang, X., Van Dyk, S.~D.,
Cuillandre, J.-C., Foley, R.~J., \& Filippenko, A.~V.\ 2007, \apj,
661, 1013

\bibitem[Ma{\'{\i}}z-Apell{\'a}niz et al.(2004)]{maiz04}
Ma{\'{\i}}z-Apell{\'a}niz, J., Bond, H.~E., Siegel, M.~H., Lipkin, Y.,
Maoz, D., Ofek, E.~O., \& Poznanski, D.\ 2004, \apjl, 615, L113

\bibitem[Mason et al.(2001)]{mason01} Mason, K.~O., et al.\ 2001,
\aap, 365, L36

\bibitem[Mattila et al.(2008)]{mat08} Mattila, S., Smartt, S.~J.,
Eldridge, J.~J., Maund, J.~R., Crockett, R.~M., \& Danziger, I.~J.\
2008, \apj 688, L91

\bibitem[Maund \& Smartt(2005)]{ms05} Maund, J.~R., \& Smartt, S.~J.\
2005, \mnras, 360, 288

\bibitem[Maund et al.(2005)]{ma05} Maund, J.~R., Smartt, S.~J., \&
Danziger, I.~J.\ 2005, \mnras, 364, L33

\bibitem[Moore et al.(2002)]{moore02} Moore, B.~D., Hester, J.~J.,
Scowen, P.~A., \& Walter, D.~K.\ 2002, \aj, 124, 3305

\bibitem[O'Connell(1999)]{ocon99} O'Connell, R.~W. 1999, \araa, 37,
603

\bibitem[Osterbrock(1989)]{osterbrock} Osterbrock, D.~E.\ 1989,
Astrophysics of Gaseous Nebulae and Active Galactic Nuclei
(Mill Valey, CA: Mill University Science Books), 21

\bibitem[Pflamm-Altenburg \& Kroupa(2007a)]{pflk07a} 
Pflamm-Altenburg, J., \& Kroupa, P.\ 2007, \mnras, 375, 855 

\bibitem[Pflamm-Altenburg \& Kroupa(2008)]{pflk07b} Pflamm-Altenburg,
J., \& Kroupa, P.\ 2008, in IAU Symposium, 246, Dynamical Evolution of Dense Stellar Systems,
ed. E. Vesperini, M. Giersz \& A. Sills (Cambridge: Cambridge University Press), 71

\bibitem[Pflamm-Altenburg et al.(2007)]{pfl07} Pflamm-Altenburg, J.,
Weidner, C., \& Kroupa, P.\ 2007, \apj, 671, 1550

\bibitem[Pilyugin et al.(2004)]{pil04} Pilyugin, L.~S.,
V{\'{\i}}lchez, J.~M., \& Contini, T.\ 2004, \aap, 425, 849

\bibitem[Poole et al.(2008)]{pool07}
Poole, T.~S., et al.\ 2008, \mnras, 383, 627

\bibitem[Pun et al.(1995)]{pun95} Pun, C.~S.~J., et al. 1995, \apjs,
99, 223

\bibitem[Ramya et al.(2007)]{ramy07} Ramya, S., Sahu, D.~K., \&
Prabhu, T.~P.\ 2007, \mnras, 832

\bibitem[Renzini \& Buzzoni(1986)]{rebuz86} Renzini, A., \& Buzzoni,
A.\ 1986, Spectral Evolution of Galaxies, 122, 195

\bibitem[Roming et al.(2005)]{rom05} Roming, P.~W.~A., et al. 2005,
Space Science Reviews, 120, 95

\bibitem[Sahu et al.(2006)]{sahu06} Sahu, D.~K., Anupama, 
G.~C., Srividya, S., \& Muneer, S.\ 2006, \mnras, 372, 1315 

\bibitem[Schlegel et al.(1998)]{sfd} Schlegel, D. J., Finkbeiner,
D. P., \& Davis, M. 1998, \apj, 500, 525

\bibitem[Sirianni et al.(2005)]{sir05} Sirianni, M., et al. 2005,
\pasp, 117, 1049

\bibitem[Skrutskie et al.(1997)]{2mass} Skrutskie, M.~F., et al.\
1997, in The Impact of Large Scale Near-IR Sky Surveys,
eds. F. Garzon et al. (Dordrecht: Kluwer Academic Publishing Company),
25

\bibitem[Smartt et al.(2008)]{smartt08} Smartt, S. J., Eldridge, J. J., 
Crockett, R. M., \& Maund, J. R. 2008, {\it MNRAS}, submitted 
(arXiv:0809.0403)

\bibitem[Van Dyk et al.(2003)]{vand03} Van Dyk, S.~D., Li, W., \&
Filippenko, A.~V.\ 2003, \pasp, 115, 1

\bibitem[V{\'a}zquez \& Leitherer(2005)]{vaz05} V{\'a}zquez, G.~A., \&
Leitherer, C.\ 2005, \apj, 621, 695

\bibitem[Vink{\'o} et al.(2006)]{v06} Vink{\'o}, J., et al.\ 2006,
\mnras, 369, 1780 (Paper~I)

\bibitem[Wang et al.(2005)]{wang05} Wang, X., Yang, Y., Zhang, T., Ma,
J., Zhou, X., Li, W., Lou, Y.-Q., \& Li, Z.\ 2005, \apjl, 626, L89

\bibitem[Woosley et al.(2002)]{woo02} Woosley, S.~E., Heger, A., \&
Weaver, T.~A.\ 2002, Reviews of Modern Physics, 74, 1015

\end{thebibliography}
\end{document}